\documentclass[journal]{vgtc}                     % final (journal style)
\onlineid{1813}

\vgtccategory{Research}

\vgtcpapertype{theory/model}

\title{On Defining Chart Type Boundaries}
\author{%
  \authororcid{Chang Han}{0000-0003-2285-1247},
  \authororcid{Andrew M. McNutt}{0000-0001-8255-4258}, and 
  \authororcid{Katherine E. Isaacs}{0000-0002-9947-928X}
}

\authorfooter{
All authors are with University of Utah. \\
Emails: \{chang.han, andrew.mcnutt\}@utah.edu, kisaacs@sci.utah.edu
}

\abstract{%
What makes a Gantt chart? This question proved unexpectedly difficult to answer when we set out to build a design space for Gantt charts.
Existing definitions, each shaped by their respective research goals, made different scope choices that we could not directly reconcile.
We reasoned about what should and should not count as a Gantt chart, developing concepts and tools along the way.
We distinguish features that are essential to a chart type's identity from those that can vary, and use these distinctions to map how chart types relate through what they share and lack.
Applying these ideas to Gantt charts, radar charts, and table cartograms, we produce key insights on what boundary work reveals: definitions diverge for functional reasons, drawing boundaries exposes hidden structure in descriptive vocabulary such as feature entanglements, and scope choices shape how far findings can generalize.
We came to understand that there is not a definitive answer, but that working through the question produced a functional definition that guided the design space we originally set out to build.
Additionally, we present vocabulary and tools for reasoning about chart type boundaries and surfacing these boundary decisions, alongside a documented Gantt chart design space.
Our broader reflection is that scope choices in chart-type-centered research---which determine what design spaces include, what grammars generate, and what perceptual studies measure---are research decisions worth making visible.

}


\graphicspath{{figs/}{figures/}{pictures/}{images/}{./}} % where to search for the images

\usepackage{tabu}                      % only used for the table example
\usepackage{booktabs}                  % only used for the table example
\usepackage{lipsum}                    % used to generate placeholder text
\usepackage{mwe}                       % used to generate placeholder figures
\usepackage{xcolor}
\usepackage{enumitem}
\usepackage{pifont}
\usepackage{wrapfig}
\usepackage{amsmath}
\usepackage{amssymb}
\usepackage{soul}
\usepackage{cuted}
\usepackage[many]{tcolorbox}
\usepackage{float}
\usepackage{graphicx}
\usepackage{calc} 
\usepackage{tikz}
\usepackage{calc}
\usetikzlibrary{calc}

\definecolor{deepgreen}{rgb}{0.0, 0.5, 0.0}
\definecolor{deepred}{rgb}{0.6,0,0}
\newcommand{\chang}[1]{{\color{black}{#1}}}

\newcommand{\essence}[1]{\ul{\textsc{#1}}}
\newcommand{\etal}{et al.}
\newcommand{\etals}{\etal{}'s}

\newcommand{\eg}{{e.g.,}}

\newcommand{\ala}{\`a la}

\newcommand{\figref}[1]{\hyperref[#1]{Fig.~\ref*{#1}}}
\newcommand{\secref}[1]{\hyperref[#1]{Sec.~\ref*{#1}}}

\newcommand{\paraheadd}[1]
{%
    \vspace{0.07in}%
    \noindent%
    \textbf{\textit{#1}}%
}
\newcommand{\parahead}[1]{\paraheadd{#1.}}

\newcommand{\numInstances}{84}

\definecolor{quoteColor}{HTML}{C46299}

\usepackage[normalem]{ulem}
\definecolor{linkColor}{HTML}{257E98}
\setuldepth{Berlin}

\newcommand{\osf}{\url{https://osf.io/j2c7k}}

\definecolor{castingColor}{HTML}{FFCC00}
\definecolor{rehearsingColor}{HTML}{AECFA4}
\definecolor{showTimeColor}{HTML}{D7B6C6}

\newtcbox{\castBox}{
  on line,
  colback=castingColor,
  colframe=castingColor,
  boxrule=0pt,
  arc=0pt,
  boxsep=1pt,
  left=0pt,
  right=0pt,
  top=1pt,
  bottom=1pt
}

\newtcbox{\rehearsingBox}[1][]{
  on line,
  colback=rehearsingColor,
  colframe=rehearsingColor,
  boxrule=0pt,
  arc=0pt,
  boxsep=1pt,
  left=0pt,
  right=0pt,
  top=1pt,
  bottom=1pt,
  #1
}

\newtcbox{\showTimeBox}{
  on line,
  colback=showTimeColor,
  colframe=showTimeColor,
  boxrule=0pt,
  arc=0pt,
  boxsep=1pt,
  left=0pt,
  right=0pt,
  top=1pt,
  bottom=1pt
}

\newcommand{\decideWord}[1]{%
  {\bgroup
    \markoverwith{%
      \tikz[baseline]{
        \fill[castingColor,fill opacity=0.85]
          (0,-0.85ex) rectangle (1.8pt,0.45ex);
      }%
    }%
    \ULon{#1}%
  }%
}
\newcommand{\formWord}[1]{%
  {\bgroup
    \markoverwith{% 
      \tikz[baseline]{
        \fill[rehearsingColor,fill opacity=0.85]
          (0,-0.85ex) rectangle (1.8pt,0.45ex);
      }%
    }%
    \ULon{#1}%
  }%
}
\newcommand{\verifyWord}[1]{%
  {\bgroup
    \markoverwith{%
      \tikz[baseline]{
        \fill[showTimeColor,fill opacity=0.85]
          (0,-0.85ex) rectangle (1.8pt,0.45ex);
      }%
    }%
    \ULon{#1}%
  }%
}
\definecolor{castingAnnotTextColor}{HTML}{8F0000}
\definecolor{castingAnnotArrowColor}{HTML}{8F0000}
\definecolor{rehearsingAnnotTextColor}{HTML}{355F2A}
\definecolor{rehearsingAnnotArrowColor}{HTML}{4A7A39}

\newtcolorbox{boxH}{
    boxrule = 0pt, 
    leftrule = 6pt % left rule weight
}

\usepackage{mathptmx}                  % use matching math font

\begin{document}

\renewcommand{\figureautorefname}{Fig.}
\renewcommand{\sectionautorefname}{Sec.}
\renewcommand{\subsectionautorefname}{Sec.}

%%%%%%%%%%%%%%%%%%%%%%%%%%%%%%%%%%%%%%%%%%%%%%%%%%%%%%%%%%%%%%%%
%%%%%%%%%%%%%%%%%%%%%% START OF THE PAPER %%%%%%%%%%%%%%%%%%%%%%
%%%%%%%%%%%%%%%%%%%%%%%%%%%%%%%%%%%%%%%%%%%%%%%%%%%%%%%%%%%%%%%%


\maketitle
\section{Introduction}

When we set out to build a design space for Gantt charts, we expected the question of ``what makes a Gantt chart?'' to have a ready answer, or at least one we could converge on quickly. This question mattered because design spaces need boundaries: they determine which examples to gather, which features to catalog, and which design alternatives to consider. But as we collected examples from academic papers and practitioner tools, the boundary proved elusive. 
% Existing definitions in the literature each worked well for their own purposes (e.g., one scope tuned for perceptual comparison~\cite{ComparativeTang} foregrounds color and order attributes, while another suited to task taxonomy~\cite{sakin2024literature} considers data attributes and task constraints). But we could not directly reconcile them for our usage. 
Existing definitions each reflected their own study's goals. For example, one scoped for perceptual comparison emphasizes fixed mark order and color~\cite{ComparativeTang}, while another for task taxonomy emphasizes event dependencies~\cite{sakin2024literature}. But we could not directly reconcile them for our usage.
What began as a practical matter of scope became a sustained investigation into how chart-type boundaries work and what decisions they carry.

That question, we came to realize, was not specific to Gantt charts. Chart types serve as organizing concepts across a wide range of visualization research: perceptual studies measure how people read specific chart forms~\cite{rensink2010perception, kong2010perceptual}, design spaces enumerate their variations~\cite{brehmer2016timelines, schottler2021visualizing, munznerDesignSpace, bertin1983semiology}, grammars and authoring tools specify how they are constructed~\cite{wickham2010layered, satyanarayan2015reactive, wang2025dataformulator2iteratively, gadhave2019upset}.

% Despite the Grammar of Graphics'~\cite{wilkinson2012grammar} argument for decomposing charts into marks, scales, and encodings, chart-type-centered discourse remains prevelant for its communicative clarity and pragmatic value as a shared organizing vocabulary.

Each of these activities depends on a scope, a commitment to what counts as an instance of the chart type under discussion. That boundary defines the object of the study: it clarifies which cases belong, which variations matter, and how a familiar label such as ``Gantt chart'' or ``word cloud'' are being used for the purpose at hand. Scope is therefore not just a matter of terminology, but part of how chart-type-centered research becomes interpretable and connectable across contexts.
To approach this question, we assembled concepts and tools to help structure our reasoning---some existing and others adapted for this context.
We developed a distinction between features essential to a chart type's identity and features that can vary across its instances, and adapted Hasse diagrams~\cite{birkhoff1948lattice, davey2002introduction} from order theory as a way to map how chart types relate through shared and absent features, turning the question of ``what makes a chart?'' into a structure that could be challenged and revised via iteration. These tools did not replace subjective judgment; they enabled us to externalize, to see the consequences of each definitional choice and to trace the reasoning behind it.

% We found it useful to distinguish features that are essential to a chart type's identity---those it must have to count as that type---from features that can vary across its instances. This distinction provided a basis for mapping how chart types relate to one another: by asking what happens when an essential feature is absent, we could systematically trace which chart types sit nearby and which are distant. We adapted Hasse diagrams~\cite{birkhoff1948lattice, davey2002introduction} from order theory to visualize these relationships, turning the intuitive query of ``what makes a chart?'' into a structure that could be challenged and revised through iteration. These tools did not replace subjective judgment; they gave us a way to externalize it---to see the consequences of each definitional choice and to trace the reasoning behind it.

We explore this question in detail through Gantt charts (\autoref{sec:case-gantts}), following an iterative process of proposing, testing, and revising candidate definitional features.
We then apply the same tools to radar charts, which surfaced complementary boundary decisions, and to table cartograms, to explore a different use: situating an unfamiliar type among neighboring charts (\autoref{sec:other-charts}).
Together, these cases show that defining chart type boundaries involves a range of decisions: which features depend on
another, at what granularity to decompose them, and how precisely to describe them.
Chart types that look different can share most of their features, such as radar charts and parallel coordinates differ only by axis arrangement, while charts that look similar or share a name can still encode data differently, as with radar charts and radial line charts.
% Chart types that look different can share most of their features, and vice versa. 
% Commonly used terms can obscure distinctions that matter for these decisions---words like ``dependencies''---which in Gantt charts alone can refer to cross-track scheduling constraints, hierarchical containment, or implicit ordering---can take on different meanings depending on context.
These contrasts made clear that chart type boundaries are not settled by visual resemblance alone, but by which structural distinctions a definition treats as consequential.
% , for example, can take on different meanings depending on context. 
The definitions we started from turned out not to be competing
attempts at the same answer, but different purpose-built definitions shaped by their respective purposes.

We did not arrive at \textbf{the} definition of Gantt chart. Over the course of the investigation, we came to understand that no such definition exists independently of the purpose it serves---a conclusion anticipated by the very definitions we started from, each functional within its own study but none universal. Chart types do not exist in nature the way molecules do, waiting to be discovered; 
% kinds like moleculars waiting to be discovered; 
they are human constructions whose boundaries are shaped by use.
The investigation produced a working definition, but, more importantly, it surfaced the boundary decisions that any such definition requires; decisions that chart-type research often leaves implicit.
% The investigation did produce a working definition---but more importantly, it surfaced the range of boundary decisions that constructing one requires, decisions that chart-type research typically leaves implicit. 
This paper traces that investigation, answering calls~\cite{akbaba2024entanglements} to make the epistemological commitments  visible.

% But the investigation produced a functional definition that proved adequate for the design space we originally set out to build.

% What a chart-type definition includes shapes the object of every study built on it---which examples belong, which variations matter, which findings transfer. 
% These downstream consequences make scope choices consequential research decisions.
\chang{From our investigation of \textit{defining chart type boundaries}}, we find that definitions diverge for functional reasons rather than by error, that drawing boundaries surfaces hidden structure in how chart types relate: feature entanglements, misleading visual similarities, and insufficient descriptive granularity, and that scope choices shape how far findings can generalize. 
We also contribute vocabulary and tools for distinguishing, describing, and mapping chart type boundaries, as well as a Gantt chart design space.
% We contribute vocabulary and tools for distinguishing, describing, and mapping chart type boundaries, longside a documented Gantt chart design space. 
% We also distill three insights about what boundary work reveals: that definitions diverge for functional reasons, that drawing boundaries surfaces hidden structure in how chart types relate, and that scope choices shape what findings can transfer.
Our broader reflection is that scope decisions in chart-type-centered research are worth making visible.

% \begin{figure*}
%     \centering
%     % \includegraphics[width=1\linewidth]{figures/pipeline.png}
%     \includegraphics[width=1\linewidth]{figures/pipeline.pdf}
%     \caption{
%     An overview of our framework and its supporting techniques: \textbf{Casting} to establish the analysis goal and curate example charts; \textbf{Rehearsing} to iteratively form and refine essences; and \textbf{Showtime} to apply the exterior design space and process artifacts to meet functional goals.
%     }
%     \label{fig:pipe}
%     \vspace{-2em}
% \end{figure*}

\section{Related Work}
\label{sec:bg-rw}

We situate our work in three areas: chart types as research objects, notions of essentialism, and set of related methodological traditions.
% our approach draws on.

\subsection{Chart Types and Design Spaces}

We discuss how chart types function as organizing concepts in visualization and how design spaces, the most closely related research tradition, rest on scoping choices that are rarely examined.

\parahead{Chart Types as Organizing Concepts}
Chart types---named visual forms such as bar charts, treemaps, and scatterplots---serve as organizing concepts across a wide range of visualization activities.
Perceptual studies measure how people read specific chart forms~\cite{rensink2010perception, kong2010perceptual}, design spaces enumerate their variations~\cite{brehmer2016timelines, tennekes2021design}, grammars specify how they are constructed~\cite{wickham2010layered, satyanarayan2015reactive}, and authoring tools implement them~\cite{wang2025dataformulator2iteratively, mcnutt2021integrated}.
Chart types also anchor practitioner-facing resources such as chart galleries~\cite{dataviz_catalogue, yang2024considering}, making them a primary vocabulary through which a great deal of visualization knowledge is communicated.

The Grammar of Graphics~\cite{wilkinson2012grammar} influentially argues for decomposing visualizations into constituent marks, scales, and encodings rather than treating chart types as atomic units.
This decomposition is powerful: it enables flexible chart construction and underlies systems such as ggplot2~\cite{wickham2010layered} and Vega-Lite~\cite{satyanarayan2015reactive}.
Yet, named chart types continue to be useful cognitive shortcuts: they chunk complex visual specifications into memorable, communicable units that practitioners and researchers can reason about without enumerating low-level encodings.

When conducting chart-type-centered research, such as building a design space or conduct perceptual studies, investigators must decide what counts as an instance of the chart type under study.
% A perceptual study of ``bar charts'' must determine whether stacked bars, grouped bars, and diverging bars all qualify.
% A design space for ``Gantt charts'' must decide whether icicle plots or temporal network diagrams belong.
% These scope commitments are consequential: they determine which examples are gathered, which features are cataloged, and whether findings can be compared across studies that use the same label.
This is a form of \textit{operationalization}~\cite{adcock2001measurement}---the translation of an abstract concept into specific, workable terms.
% When different studies operationalize the same chart-type name differently, their findings may not be directly comparable, an instance of what has been called the \textit{jingle fallacy}~\cite{kelley1927interpretation}: same label, different meanings. For instance, ``heatmap'' can denote both grid-based matrix visualizations and geographic point-density surfaces~\cite{deboer2015understanding}---structurally distinct designs whose shared label can obscure that findings about one need not apply to the other.
In chart-type-centered research, operationalization means deciding which visual designs count as instances of the type under study, for example, whether ``bar charts'' includes stacked, grouped, or diverging variants.
% For example, the term ``radar chart'' is used for both true radar charts (which assign discrete, independent axes to separate data dimensions) and radial line charts (which use a continuous angular dimension).
% These encode data quite differently, yet share a name; a perceptual study of one may not generalize to the other.
% Explicit operationalization is an established methodological principle in the social sciences; in visualization research, scope choices for chart types are rarely made explicit and are typically settled by intuition.

\parahead{Design Spaces and Chart Modeling}
Among prior work that centers specific chart types, our work is closely related to design spaces.
Design spaces in visualization range from specific chart types like timelines~\cite{brehmer2016timelines} to visualizations in particular domains~\cite{crisan2018method} or with certain data types~\cite{tennekes2021design}.
Munzner~\cite{munznerDesignSpace} reviews design spaces in visualization and describes lessons learned from developing several.
Tennekes and Chen~\cite{tennekes2021design} connect design spaces with related models of knowledge including taxonomy and ontology, arguing that design spaces facilitate comparison by listing alternatives as parameters.
Design spaces are also a subject of study in software engineering~\cite{shaw2011role, Shaw2024Design}, where Shaw argues they help organize design alternatives and make decision rationale explicit.
% Sedlmair \etals{}~\cite{sedlmair2012design} design study methodology offers a structural analog to our work: a methodological contribution that emerged from doing the work and reflecting on the process.
Sedlmair \etals{}~\cite{sedlmair2012design} design study methodology, similarly to our work, emerged from reflecting on repeated practice, distilling structure from experience conducting design studies.

\chang{Visualization taxonomy work also involves scoping commitments. Chen \etal{}~\cite{chen2022not} describe how developing a bottom-up typology of visualization images required resolving ambiguity and repeated refinements to category definitions. VisTaxa~\cite{zhang2025vistaxa} similarly treats taxonomy development as an empirical coding protocol, including decisions about definition formatting and boundary cases such as the distinction between visualizations and maps.}

% Notably, each of these design spaces rests on scoping choices.
% Brehmer \etal{}~\cite{brehmer2016timelines} derive their timeline design space from an iteratively assembled corpus of examples, while Crisan \etal{}~\cite{crisan2018method} 
% % 
% develop a design space for genomics visualizations by surveying domain literature, sampling existing tools, and coding their features.
% through systematic survey---but the scoping decisions themselves are typically addressed in passing rather than examined as a subject in their own right.
% Our work complements design space construction by making this scoping step explicit. Where design spaces typically focus inward by enumerating what a chart type contains and how its instances vary, our work focuses on the boundary itself: what belongs in the space, what does not, and how those decisions relate to neighboring chart types.

\chang{
Prior work has likewise treated named chart types as members of broader visualization families.
Wickham and Hofmann's product plots~\cite{wickham2011product} show that many area-based visualizations, such as bar charts and treemaps, can be described through shared constraints and recursive partitioning primitives.
Schulz and Hadlak~\cite{schulz2015preset} extend this view through parametric visualization designs and preset-based blending, allowing designers to interpolate between existing techniques; they similarly constructed a Hasse diagram for product plots, to illustrate how feature combinations can organize neighboring designs.
Li \etal{}~\cite{li2015exploring} use feature vectors and phylogenetic trees to model visualization designs.
% These works suggest that named chart types are not isolated categories: their boundaries are dynamic, their features can be entangled, and neighboring designs can differ by varying degrees.
Together, these works show that feature choice, family resemblance, and neighbor relations are recurring concerns in visualization design spaces rather than peculiarities of Gantt charts.
% These works show that feature choice, family resemblance, and neighbor relations are recurring concerns in visualization design spaces.

Our work complements the design space and chart modeling work by making the scoping step explicit. Where design spaces often organize or generate chart families after relevant features have been selected, our work focuses on the boundary itself: what belongs in the space, what does not, and how those boundaries relate to neighboring chart types.}
\subsection{Concepts and Definitions}

We discuss prior work that examines foundational concepts in Visualization and HCI, then introduce the philosophical traditions we draw on to reason about chart-type definitions.

\parahead{Definitional and Conceptual Work}
The Visualization and HCI research communities have (re)examined some familiar concepts in the field.
Hornb\ae{}k and Oulasvirta~\cite{hornbaek2017interaction} ask ``What is interaction?'', a concept central to HCI yet lacking a settled definition, and show that examining it carefully reveals hidden assumptions and productive distinctions.
Beaudouin-Lafon~\etal{}~\cite{beaudouin2004designing, beaudouin2021generative} similarly interrogate foundational concepts in interaction design, developing theoretical frameworks that reframe how the field understands its own terms.
Bowker and Star~\cite{bowker2000sorting} study how classification systems shape what is visible and what is tacit, arguing that the categories we use have real consequences for practice.
Brehmer and Munzner~\cite{brehmer2013multi} address fragmented characterizations of visualization tasks by constructing a multi-level typology that distinguishes tasks as \textit{why}, \textit{how}, and \textit{what}, showing how principled structure applied to a concept space enables richer analysis and cross-domain comparison.
Our paper joins this conversation: chart-type names are categories that visualization research relies on, and examining their boundaries reveals decisions worth making visible.

Burns et al.~\cite{burns2023we} document a parallel case for audience categories: the label ``novice'' is used broadly, yet papers rarely make explicit what the term means or who it includes, leading to narrow operationalizations that are difficult to compare or generalize. Our argument asks for the same explicitness about chart-type names.

\parahead{Essentialism and Chart Types}
To reason about scope, we draw on the philosophical tradition of essentialism~\cite{sep-essential-accidental}, which distinguishes between properties an object \textit{must} have (essential properties) and properties it \textit{could} have but does not need (accidental properties).
We adapt this distinction as a lens for chart types: which features must a chart have to count as a given type (essential features), and which can vary across instances (variable features)?
For example, a bar chart must use length along a common axis to encode quantitative values, this is essential to its identity.
But the specific color of the bars, their orientation, or whether they are sorted is variable: changing these does not make it stop being a bar chart.
% \chang{Note that we do not take a strong essentialist position about chart types.}

\chang{Essentialism has long been the subject of philosophical debate~\cite{sep-essential-accidental}, and fully engaging that debate is beyond our scope; nor do we take a strong essentialist position. We make no claim that chart types have context-independent essences; rather, we use the essential–variable distinction pragmatically to structure boundary decisions, as we further discuss in~\autoref{sec:disco}.}

\subsection{Reasoning Frameworks and Methods}

We discuss methodological traditions our work intersects: structured
reasoning frameworks, qualitative methods, and sensemaking models.

\parahead{Structured Reasoning in HCI and Visualization}
Our work aligns with structured reasoning approaches such as QOC~\cite{maclean2020questions} by externalizing definitional and scoping choices, making assumptions about what counts as an instance visible for comparison, reflection, and critique.
% Our work shares an orientation with approaches that provide structured reasoning aids for design and research decisions.
% MacLean \etals{}~\cite{maclean2020questions} Questions, Options, and Criteria (QOC) notation offers a structured format for reasoning about design alternatives and making the rationale behind choices explicit.
% Just as QOC externalizes the space of design alternatives so that rationale can be inspected and challenged, our exploration of structured scope reasoning externalizes the space of definitional choices---which features are essential, which are variable, and what the consequences of each choice are.
% Our tools serve an analogous function for a different kind of decision: not which design alternative to choose, but what counts as an instance of the design in the first place.
Meyer and Dykes~\cite{meyer2019criteria} propose six criteria for rigor in visualization design study, including transparent reporting of activities, evidence, and analysis, thereby enabling readers to scrutinize how knowledge is constructed. In a related spirit, our work likewise seeks to make an often-implicit aspect of research practice more explicit. Though we do not offer prescriptive criteria, we treat structured scope reasoning as an exploratory and reflective practice for surfacing assumptions and revealing definitional distinctions.

\parahead{Sensemaking Models and Qualitative Methodology}
Our work also connects to sensemaking models~\cite{berret2024iceberg, grolemund2014cognitive, sacha2014knowledge} in that it provides conceptual scaffolding for a specific set of sensemaking tasks.
% ---those relating to chart type definition.
Among these, Berret and Munzner's Iceberg model~\cite{berret2024iceberg} bears structural similarity, with its iterative cycles of adding material and refining understanding.
Indeed, the iterative definitional process we describe produces something akin to a schema, a structured framework for understanding what a chart type is, but our focus is on the boundary decisions within that framework: the boundary choices about what counts as an instance, which propagate into downstream research.
Where sensemaking models address general data analysis, our tools target the specific task of reasoning about what a chart type means in service of a functional goal.

The iterative process of proposing, testing, and refining essential features that we describe draws on established qualitative methods.
Braun and Clarke's~\cite{braun2006using} thematic analysis provides a foundation for coding examples and identifying patterns.
We extend this descriptive orientation with definitional reasoning by asking not only which themes recur across examples, but also \textit{which features are necessary for the concept to hold}.

\newcommand{\radarRef}[1]{\textbf{R#1}} 

\begin{figure}
  \centering
  \includegraphics[alt={Three Gantt-chart examples: a conventional chart with duration bars and dependency arrows across tracks; four spring-based schedules using stepped and zigzag lines; and a production schedule with four order rows and color-coded process bars along time.},width=1\linewidth]{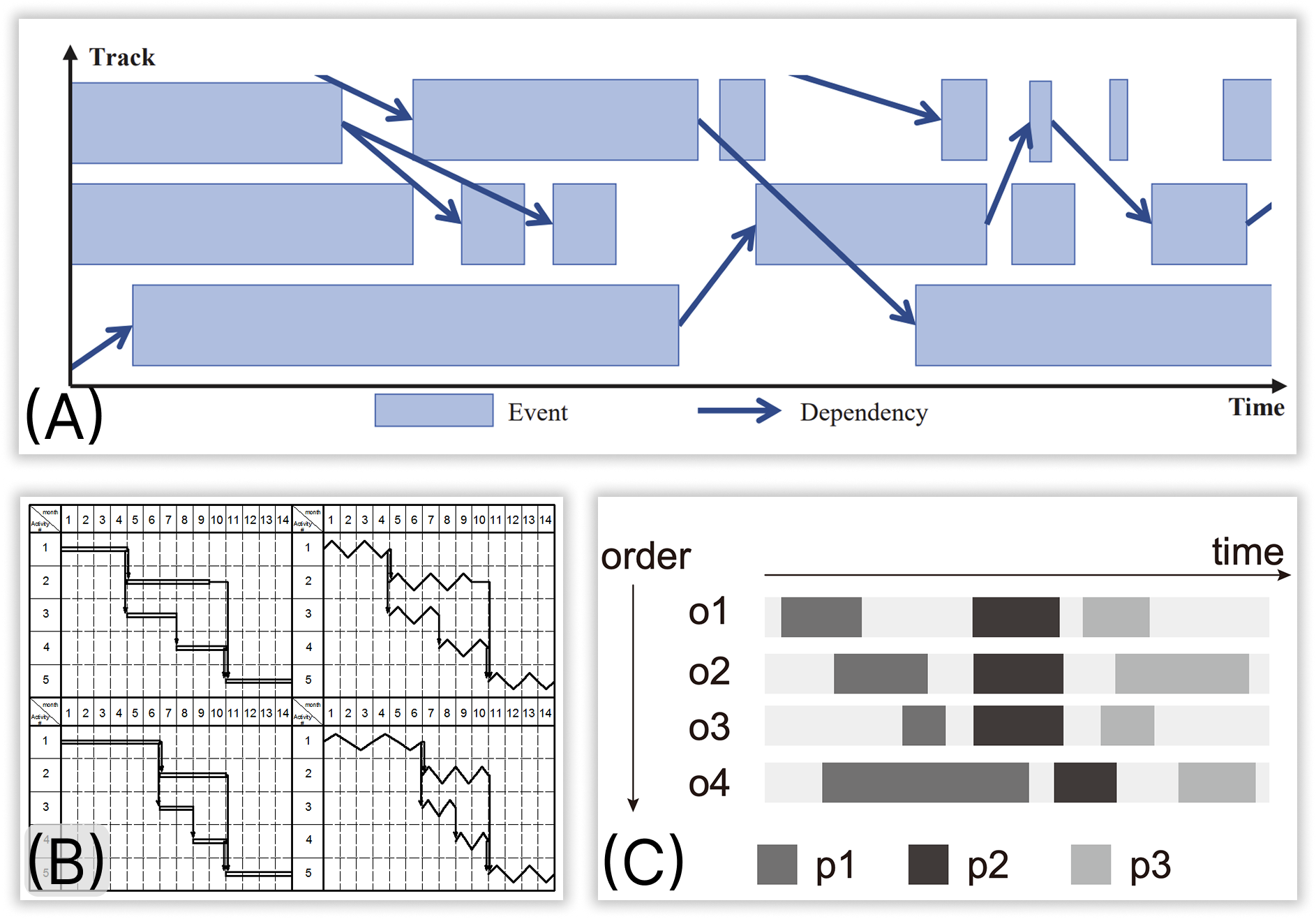}
  \caption{
    Examples from our Gantt-chart collection, (A) showing
    events with explicit arrows~\cite{sakin2024literature}, (B) modeling schedule adjustments as the compression and stretching of a spring (represented by triangle waves)~\cite{undozerov2021spring},
    and (C) a bar-based production schedule organized by order and time with categorical color encoding~\cite{tang2021visualization}.
    Their contrast in marks and relational structure illustrates the challenge of forming a single definition.
  }
  \label{fig:gantt-gallery}
  \vspace{-1.5em}
\end{figure}

\section{What Makes a Gantt Chart?}
\label{sec:case-gantts}

% Three visualization researchers engaged in weekly discussions to establish a practical design space for Gantt charts, with the intent to develop a visualization grammar for them. One additional researcher engaged in early phases of discussion.
A design space needs a boundary: it depends on a workable answer to what belongs and what does not.
On the way to build a design space for Gantt charts, a critical question we encountered was ``Just what is a Gantt chart?''
This section traces how we worked through that question---what we started from~(\autoref{sec:definitions-started-from}), how we come up with our own definitional structure~(\autoref{sec:what-must}), how we refined the definition iteratively~(\autoref{sec:probing} and \autoref{sec:mapping}), what we arrived at~(\autoref{sec:definition}), and how we applied to definition to build the design space~(\autoref{sec:showtime-case}). Across this process, we gathered \numInstances{} examples from academic papers and practitioner tools (we provide a complete list in supplemental materials), and used them to propose, test, and revise candidate features. The collection began with examples from papers and practitioner tools and was expanded iteratively as additional cases emerged during the exploration. It was also informed by candidate chart forms and examples suggested by a co-author with over a decade of experience working with Gantt charts.
Along the way, we assembled a set of techniques, some drawn from existing methods, others adapted for this context, that we use to form and assess features used for definition. 
% We provide an overview of the process in~\autoref{fig:overview}, and present details in this section.

\subsection{The definitions we started from}
\label{sec:definitions-started-from} 

We started by revisiting how prior work defines Gantt charts. Several recent works define or implicitly characterize Gantt charts. We expected these accounts to converge on a workable scope for our design space, but they did not.

The following examples illustrate this divergence: each study scopes ``Gantt chart'' differently in ways that reflect its particular research goals.
Tang \etal{}~\cite{ComparativeTang}, in a comparative perceptual study, define Gantt charts as: visual representations using fixed orders of rectangles or lines to denote events, with categorical color encoding for event types (e.g., \autoref{fig:gantt-gallery} (C)). This scope is tuned for perceptual comparison, where the visual form of marks matters directly.
Sakin and Isaacs~\cite{sakin2024literature}, constructing a task taxonomy, take a different view: their Gantt charts feature dependencies between events rather than fixed ordering (e.g., \autoref{fig:gantt-gallery} (A))---a choice that meaningfully extends the task space they can describe.
Jo et al.~\cite{Jo2014}, building interactive Gantt charts for manufacturing scheduling, implicitly adopt yet another scope: their drag-to-reschedule interaction implies constraints between bars, and their use of aggregated bars suggests hierarchical data structure.

These differences have practical consequences. A chart with dependency arrows between tracks would fall outside Tang \etals{} definition but is central to Sakin and Isaacs'. The same visualization can be included or excluded depending on which definition is adopted---and these scope choices propagate into what a design space contains, what features are cataloged, and what findings are produced.

This divergence, we came to realize, was not an oversight by prior authors. It reflected something about how chart-type boundaries work in practice: they are not fixed properties of a visual form but functional commitments shaped by what the definition is being used to do. The same chart-type name can carry different scopes across different studies, and those differences are rarely surfaced or examined. 
We initially hoped that one sufficiently broad definition of Gantt charts could serve as a common foundation. But as we assembled examples, 
we found that maximal coverage was not the right goal for our purposes.
Instead, we needed a scope whose cases shared enough design vocabulary for feature comparisons to remain meaningful.
% A design space is useful only when the cases it includes share a sufficiently common design vocabulary: it must be broad enough to expose meaningful variation, but not so broad that the included forms stop informing the same set of design choices. 
Some prior definitions were too narrow for that purpose, excluding cases whose variation we wanted to capture; a maximally broad definition would have been too loose, collapsing together forms whose features did not transfer meaningfully. The task, then, was not to recover the one true scope of ``Gantt chart,'' but to construct a scope suited to one coherent design space.
% This meant no existing definition could serve as a ready-made foundation---we needed to construct one whose scope was explicitly aligned with the goal of building a comprehensive design space.
% We ourselves also had an intuitive assumption of what a Gantt chart is---bars on a time line, organized by tracks---but found this intuition, when pressed to decide concrete inclusion cases, was neither as shared nor as precise as we had assumed.

Each definition was functional within its own study, none was wrong. But each would accommodate different chart instances.
Even the three examples in Fig.~\ref{fig:gantt-gallery} already pull the boundary in different directions:
\autoref{fig:gantt-gallery}~(A)
foregrounds dependencies between tracks,
\autoref{fig:gantt-gallery}~(B) departs from the familiar bar-based appearance through a spring notation to model schedule adjustments,
and \autoref{fig:gantt-gallery}~(C) presents a bar-based production schedule with categorical color encoding.
Together, they made clear that no single prior definition cleanly captured the range of cases we wanted to reason about.
% What followed was a sustained process of proposing, testing, and revising candidate features---one that changed not only our definition of Gantt charts but how we understood what it means to define a chart type.
We then iteratively proposed, tested, and revised candidate features, refining our definition of Gantt charts and clarifying our understanding of chart-type definition.
% Each definition was functional within its own study---none was wrong. But each is different and would accommodate different chart instances.
% Throughout the process, we gathered \numInstances{} examples from academic papers and practitioner tools (complete list in supplemental materials), and set out to build our own definition from the ground up.
% % because our goal of building a comprehensive design space did not align cleanly with any single prior scope. 
% For example,
% \textbf{G3} features dependency arrows between tracks,
% which does not fall into Tang et al.'s definition but Sakin and Isaacs'
% would require; \textbf{G1} uses non-rectangular marks (springs),
% testing whether visual form should be part of the definition at all.
% What followed was a sustained process of proposing, testing, and revising candidate features---one that changed not only our definition of Gantt charts but how we understood what it means to define a chart type.

\subsection{What must a Gantt chart have? Separating essential from variable features.}
\label{sec:what-must}

When we conducted open coding~\cite{Braun2019} on the curated examples, the list grew quickly: bar color, mark shape, timeline orientation, discrete tracks, dependency arrows, among others.
But these features were clearly not all playing the same role.
Some felt definitional---remove the timeline axis from a Gantt chart and it stops being a Gantt chart.
Others felt like design choices---\chang{remove the border thickness variation among the bars} and it is still recognizably a Gantt chart, just a less informative one.
We needed a principled way to separate these features.

% \parahead{Essential and variable features}
Drawing on essentialism (\autoref{sec:bg-rw}), we call a feature \textit{essential} if removing it changes what the chart fundamentally is, and \textit{variable} if the chart remains recognizable without it.
The practical test is an algebraic-inspired~\cite{mcnutt2021table, kindlmann2014algebraic} removal criterion: if a feature can be removed and the result is still understandable as that chart type, the feature is variable; if removing it fundamentally changes what the chart is, it is essential.
For Gantt charts, this test quickly sorted some features: removing color encoding leaves a Gantt chart that is still a Gantt chart; removing the timeline axis yields something that is not. 
But the test is less mechanical than it sounds. The answer to ``is it still a Gantt chart after removing this feature?'' is itself a subjective judgment, one that different researchers might answer differently based on their prior experience and prototypical examples. In our case, three researchers had to reach agreement on each call, and these agreements were themselves scope decisions: deliberate commitments rather than discovered facts. The removal criterion does not eliminate subjectivity; it structures it, by giving us a specific question to disagree about.
\chang{The reasoning also naturally extends to ``what-if'' checks that allow us to discuss what would happen if a judgment went the other way. For example, treating categorical colors as essential would exclude charts like~\autoref{fig:gantt-gallery}~(A).}

These iterative discussions drew on expert review, \ala{} heuristic evaluation~\cite{Tory2005, wall2018heuristic} or critical reflection~\cite{satyanarayan2019critical}, where prior knowledge and experience are used to propose and critique essences, alongside the coding.
One co-author has over a decade of experience with Gantt charts; the others brought visualization design and theory expertise.

% \ganttVoneAnnotatedFigure{figures/ganttV1.png}
% \vspace{-3mm}
% \ganttParagraphStartAnchor With this framing in hand, the first author then conducted \ganttOpenCodingText, and the group refined the results through \ganttDiscussionText, arriving at an initial set of \ganttCandidateText{} (\ganttShownText{} in the above figure). These candidates emerged from both the patterns visible in the coded examples and the domain knowledge contributed through expert review.
With this framing in hand, the first author then conducted the open coding, and the group refined the results through expert discussion, arriving at an \textbf{initial set of four candidates}:
\begin{enumerate}[nosep, leftmargin=*]
    \item \essence{Rectangles / lines encode duration}
    \item \essence{Timeline axis}
    \item \essence{Tracks (Rows) present}
    \item \essence{Dependency between marks}
\end{enumerate}
% These candidates emerged from both the patterns visible in the coded examples and the domain knowledge contributed through expert review.
% \ganttVoneOverlayArrows
% What followed was an iterative process of testing and refining these candidates, through counterexamples, linguistic scrutiny, and structural analysis.
% ---that we describe in the following subsections.

\subsection{Probing the boundaries}
\label{sec:probing}

With candidate essences in hand, we began testing them against our collected examples.
% As we assembled and expanded the collection, we deliberately included edge cases---charts that sit on the boundary of what might or might not count as a Gantt chart---alongside clear-cut positive examples. 
As we assembled and expanded \chang{(\ala{} purposeful sampling~\cite{palinkas2015purposeful})} the collection, we deliberately included edge cases, such as spring-based Gantts and icicle plots, alongside clear-cut positive examples, such as conventional bar-based Gantt charts.
Some of these edge cases turned out to be \textit{counterexamples}: charts that share surface features with Gantt charts but are not, on reflection, members of the family.
Rather than discarding them, we retained these counterexamples because they expose unexamined assumptions and provide useful friction when testing candidate essences.

In practice, probing and refining essences involved repeated revision, reconsideration, and false starts; here we highlight a few episodes that illustrate the kinds of revisions it produced.

\parahead{Generalizing mark descriptions}
Early on, we described Gantt chart marks as \textit{rectangles or lines}, following the visual forms most common in practice. But we encountered valid Gantt charts using alternative representations, such as springs~\cite{undozerov2021spring} ((B) in~\autoref{fig:gantt-gallery}), which broke this characterization without changing the chart's identity. We generalized to \textit{data marks}, which is a broader term that captured the essential role (encoding scheduled events) without over-committing to a specific visual form.

\parahead{Refining duration encoding and track specifications}
With that revision, we had \essence{data marks encode duration} as a candidate essential feature. But one example from Isaacs \etal{}~\cite{isaacs2015ordering} 
\begin{wrapfigure}[6]{ri}{0.6\linewidth}
  \vspace{-0.4cm}
    \centering
    \includegraphics[alt={A 16-row time grid with colored rectangular cells occupying single time units and black connector lines linking cells across rows. The horizontal axis runs from Lateness to 57 nanoseconds.},width=1\linewidth]{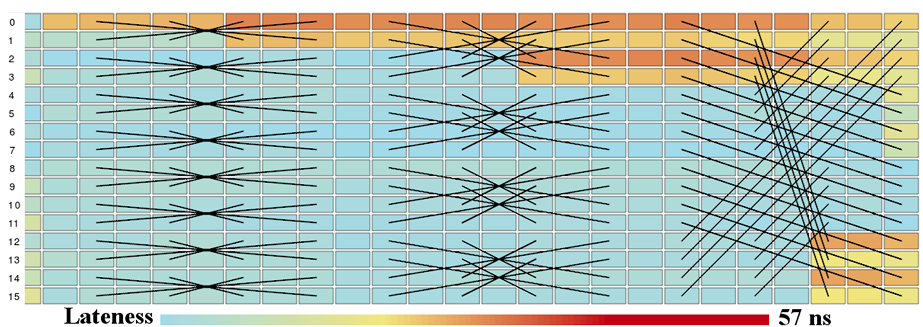}
  \vspace{-0.9cm}
\end{wrapfigure}
used uniform-length bars where all events occupied a single time unit (right).
Was this still a Gantt chart? We thought so---the bars were clearly \textit{meant} to represent time spans, even though they happened to be uniform. We refined the feature to \essence{data marks \textbf{can} encode duration}: the marks must have the \textit{capacity} for variable length, even if a particular instance does not use it. \chang{In this sense, duration encoding requires a semantic mapping between mark extent and temporal duration, not necessarily visible length variation in every displayed instance.}
This kept uniform-duration Gantt charts within scope while excluding temporal network diagrams, where nodes cannot meaningfully encode length.
We had considered broader mark types as well, but when marks could not meaningfully lengthen or shorten with duration, those representations were better understood as neighboring forms, such as node-link diagrams or dynamic graph visualizations, than as Gantt charts.

Another counterexample exposed an ambiguity in the vertical layout.
We initially described the feature simply as \essence{tracks present}.
\begin{wrapfigure}[7]{ri}{0.6\linewidth}
  \vspace{-0.4cm}
    \centering
    \includegraphics[alt={A scheduling view from 17:35 to 18:00 in which hundreds of thin job bars form continuous stacked bands that expand, contract, and split vertically rather than remaining in fixed rows.},width=1\linewidth]{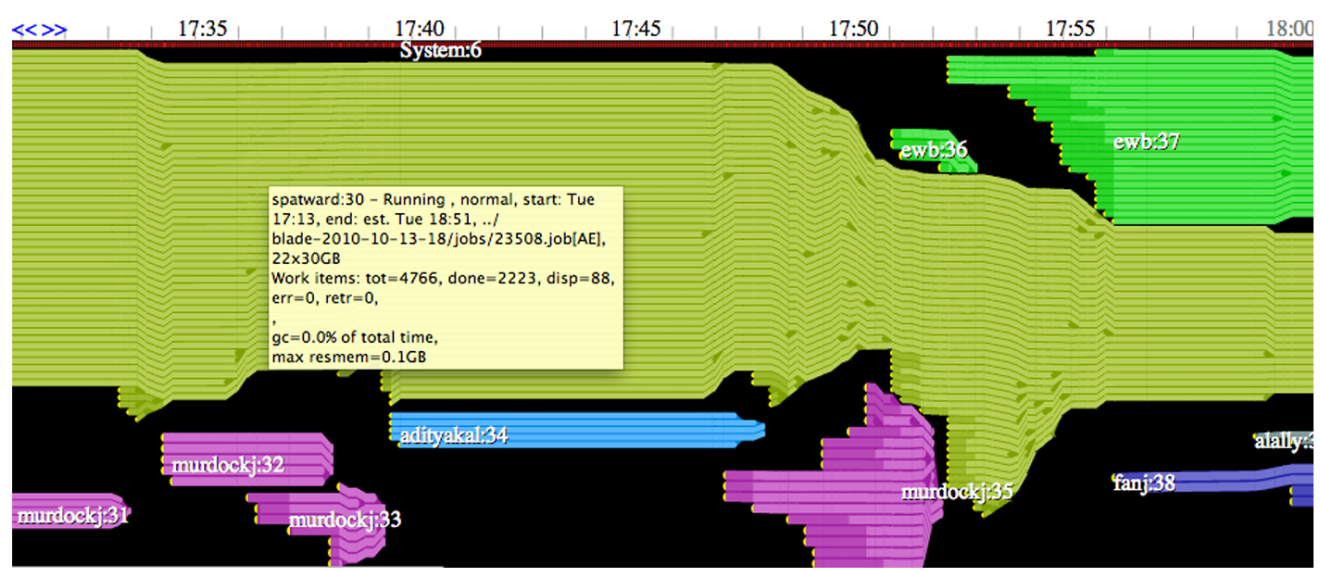}
  \vspace{-0.9cm}
\end{wrapfigure}
But one chart~\cite{de2013visualizing} (right) placed events against a y-axis representing something like resource quantity, and the bars flowed across that axis more like layers in a streamgraph than items assigned to fixed rows. 
We did not regard this as a Gantt chart: although the chart had a track-like vertical structure, its positions were continuous rather than stable lanes. To distinguish such cases more clearly, we revised \essence{tracks present} to \essence{discrete tracks}.

% A harder case arose with duration. With the above revision we have \essence{Data marks encode duration} as an essential feature. But one example from Isaacs \etal{}~\cite{Isaacs2014} used uniform-length bars where all events occupied a single time unit. Was this still a Gantt chart? We thought so---the bars were clearly \textit{meant} to represent time spans, even though they happened to be uniform. To accommodate this, we refined the essence to \essence{Data marks \textbf{can} encode duration}: the marks must have the \textit{capacity} for variable length, even if a particular instance does not use it. This excluded temporal network diagrams, where nodes cannot meaningfully encode length, while keeping uniform-duration Gantt charts within scope.
% \vspace{-5mm}
% \inlinefig{figures/ganttV2.png}
% \vspace{-5mm}

% \parahead{Splitting dependencies}
\parahead{Resolving concept collisions}
In many Gantt charts, arrows or other connectors are used to show that one depends on another, for example, ``task B cannot start until task A finishes.''
Our initial formulation treated these dependencies as a single feature.
But when we considered icicle plots or flame graphs (below), 
\begin{wrapfigure}[7]{ri}{0.3\linewidth}
  \vspace{-0.3cm}
    \centering
    \includegraphics[alt={A flame graph of Apache Cassandra call stacks. Nested green rectangles rise from wide base calls into narrower columns, with yellow and red blocks at the tallest peaks, showing hierarchical containment.},width=1\linewidth]{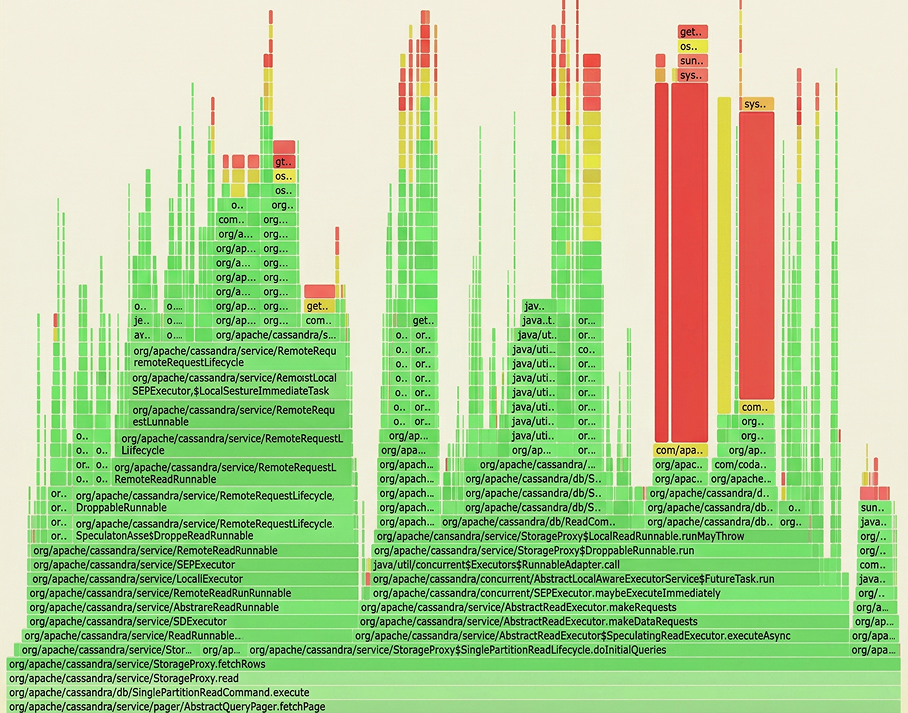}
  \vspace{-0.9cm}
\end{wrapfigure}
which also show hierarchical relationships progressing from top to bottom, we realized that not all dependencies are the same. In Gantt charts, dependencies indicate \textit{cross-track} constraints between events on different lanes; in icicle plots, they represent hierarchical containment within a tree structure. We introduced the term \textit{cross-track dependencies} to capture this distinction, making the feature more precise.
% One consequential revision concerned dependencies. Our initial formulation treated dependencies as a single feature. But when we considered icicle plots---which also feature hierarchical dependencies progressing from top to bottom---we realized that not all dependencies are the same. Those in Gantt charts indicate \textit{cross-track} constraints between events (``task B cannot start until task A finishes''), whereas those in icicle plots represent hierarchical containment within a tree structure. We introduced the term \textit{cross-track dependencies} to capture this distinction, splitting a single feature into a more precise one.

\begin{figure}[t]
    \centering
  \includegraphics[alt={Two-axis conceptual diagram for the flexibility check. The vertical axis, Visual Designs, runs from not specified at the bottom through partially specified to fully specified at the top; the horizontal axis, Data Abstractions, runs through the same levels from left to right. Example phrases occupy different regions: Circular visualizations is near the lower left; Regular convex polygon is high but left; Multivariate data is low but right; and A fully designed concrete instance of a chart is near the upper right. Lines connecting marks lies near the center. A diagonal arrow moves down and right from Lines connecting marks to Dependency between marks, showing that the revision removes visual-form specificity while adding data-relationship specificity. The positions are qualitative comparisons rather than measurements.},width=1.02\linewidth]{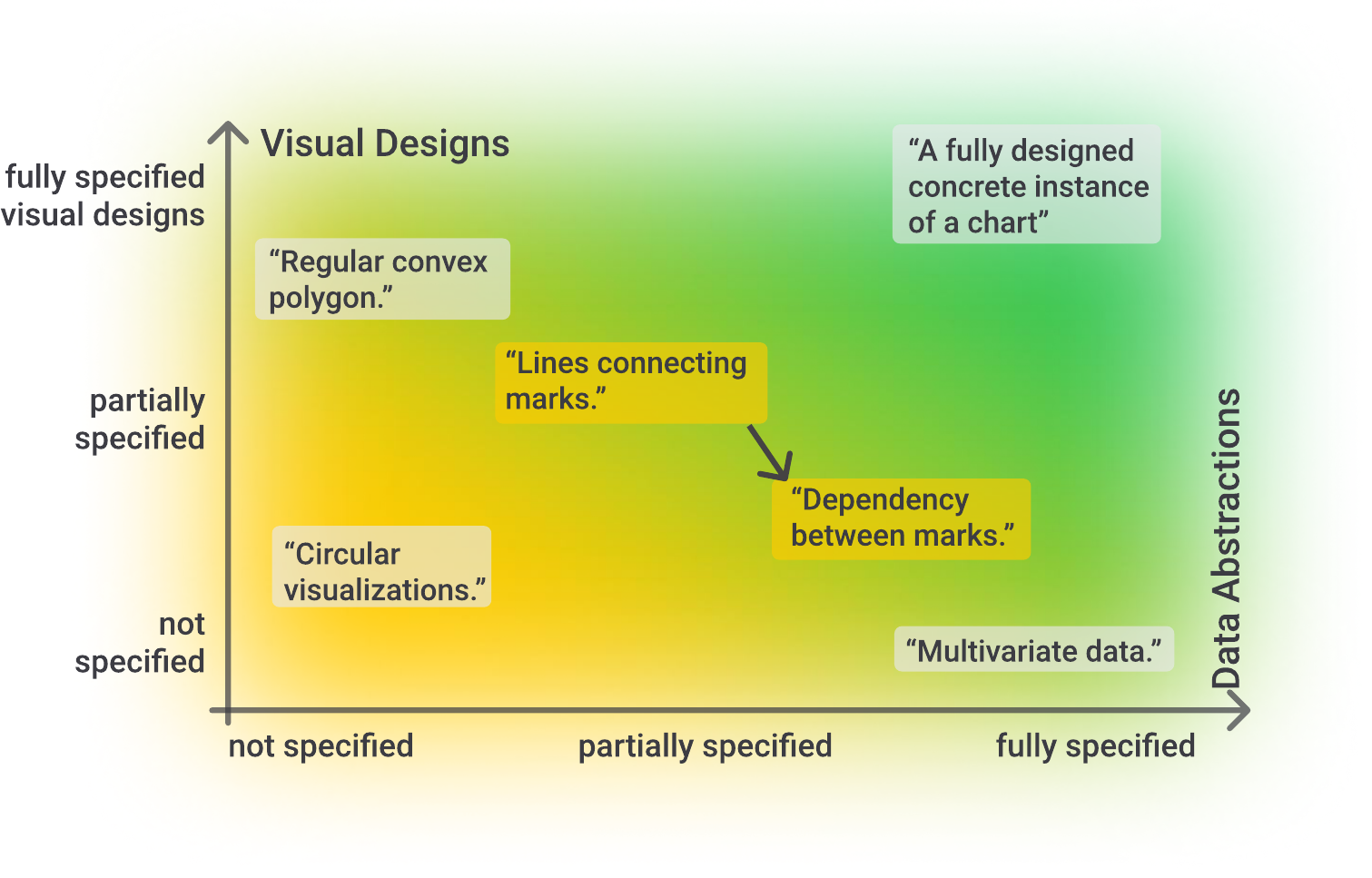}
    \vspace{-8mm}
    \caption{The flexibility check examines each essence along two axes---visual design and data abstraction---to reveal how much latitude a given phrasing permits. The arrow traces how a Gantt dependency essence shifted from a visually constrained phrasing to a data-focused one.}
    \label{fig:ClassifyingDesc}
    \vspace{-1.5em}
    % \vspace{0em}
\end{figure}

\parahead{When the words themselves are the problem}
Counterexamples test whether the right charts fall inside and outside the definition. But we discovered a different kind of issue: even an essence that survived counterexample testing could harbor unintended ambiguity in its \textit{language}.
Features are expressed in natural language, and natural language is inherently ambiguous---but not uniformly so. Different word choices govern different degrees of flexibility when describing chart features: ``regular convex polygon'' pins down a visual form while leaving the data entirely open; ``multivariate data'' constrains the data while saying nothing about appearance. A description that seems clear to its author can quietly admit interpretations they never intended.

To handle these latent ambiguities, we treated each candidate essential features as a small design exercise that we came to call a \textit{flexibility check} (\autoref{fig:ClassifyingDesc}). For each phrasing, we asked two simple questions: how much does it lock down the visual form, and how much does it lock down the underlying data abstraction? At one extreme, ``a fully designed concrete instance of a chart'' 
leave little flexibility on either axis; at the other, ``circular visualizations'' 
commit only to a visual motif. Most useful essences lie somewhere between those poles. The positions in \autoref{fig:ClassifyingDesc} should therefore be read as comparative sketches, not precise measurements.
% To surface these latent ambiguities, we developed a \verifyWord{flexibility check} (\autoref{fig:ClassifyingDesc}): a systematic examination of each essence along two axes---\textit{visual design} and \textit{data abstraction}---asking how tightly the current wording constrains each. Corner cases anchor the space: ``specific chart instances'' fully constrain both axes, while phrases like ``circular visualizations'' specify only a visual motif, leaving data entirely open. Most useful essential features sit somewhere in between, and the check forces a deliberate choice about where. Note that positions in this space shown in~\autoref{fig:ClassifyingDesc} are approximate and comparative---the check is a qualitative tool for exposing trade-offs in a phrasing, not a precise instrument for measuring.

That wording exercise directly reshaped our Gantt essences. For dependencies, we initially wrote ``lines connecting marks,'' which quietly privileged one visual solution: a dependency shown as a connector. But some of the examples we collected expressed the same dependency without any line at all, simply by placing one bar after another in sequence. Rephrasing the essence as \essence{dependency between marks} kept the relationship and released the visual form, which better matched what we considered essential.

The check does not enforce a particular level of precision; broader or narrower phrasings can both be appropriate. What matters is that the choice is deliberate rather than an accident of wording.

\begin{figure}
    \centering
  \includegraphics[alt={Hasse diagram organizing chart types by three essential Gantt features: duration encoded along timelines, discrete track axes, and cross-track dependencies. A blue check means a feature is present, a red cross means it is absent, and moving downward along one edge removes one feature. The top node, Gantt charts, has all three. Its immediate neighbors each omit one feature: temporal networks omit duration encoding; Timenets omit track axes; and lifelines or icicle plots omit cross-track dependencies. Nodes farther down retain only cross-track dependencies, exemplified by node-link diagrams; only duration encoding, exemplified by line charts; or only track axes, exemplified by bar charts. The bottom node has none and represents everything else. The structure shows which chart types are one feature away from Gantt charts and which are more distant.},width=1\linewidth]{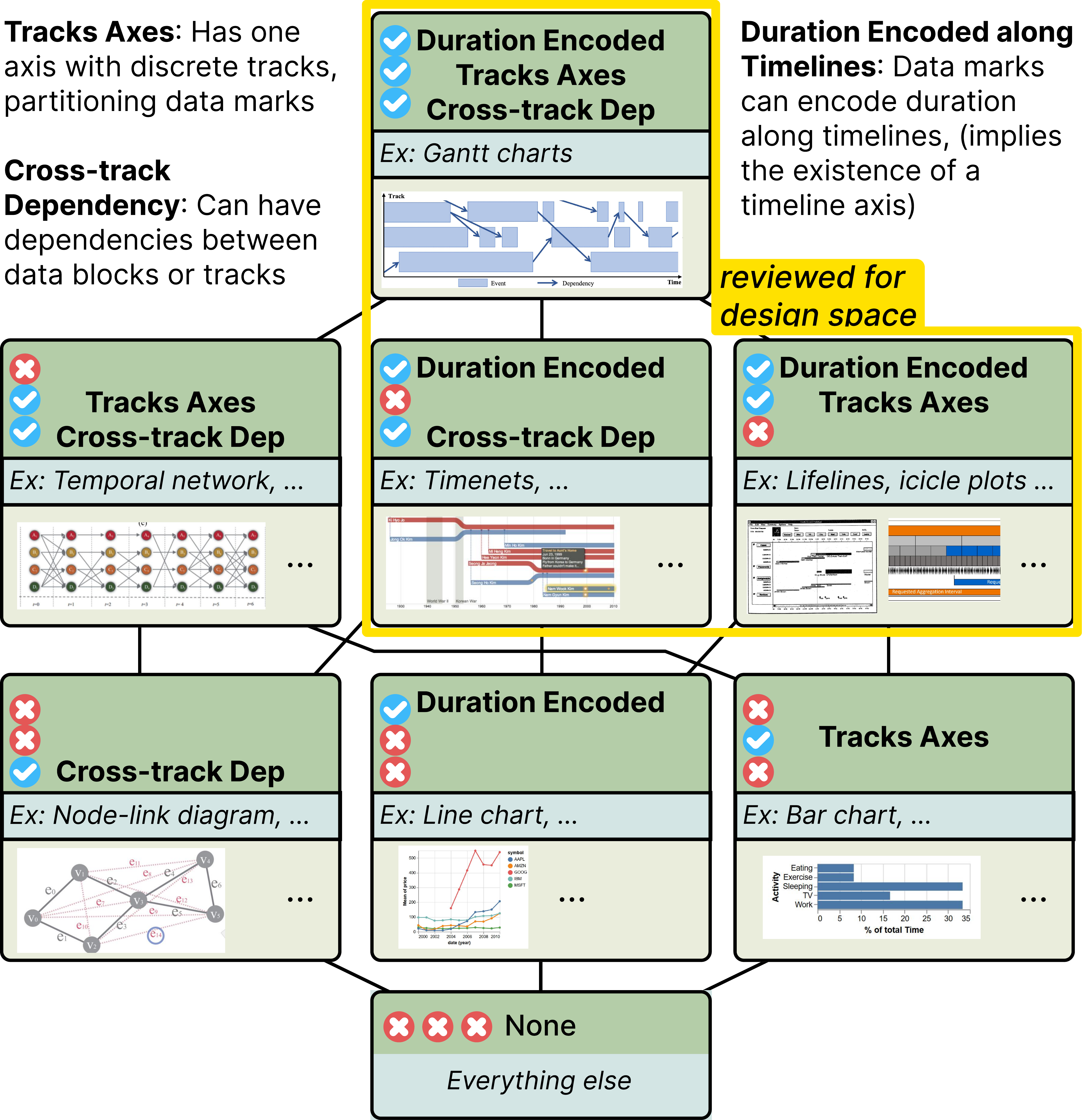}
    \caption{
    The Hasse diagram for Gantt charts from our analysis. In addition to the root node (Gantt charts), two other nodes were reviewed for developing the design space.
    }
    \label{fig:gantt-hasse}
    \vspace{-1.5em}
\end{figure}

\subsection{Mapping chart-type relationships}
\label{sec:mapping}

The revisions above refined individual features. But once we had a candidate set of essential features, a natural question arose: how do these features interact with one another, and what chart types sit nearby that share some features but not all?

% Not every chart satisfies all of our essential features. Some share two out of three, placing them close to Gantt charts without being one. Others share only one, making them more distant relatives. We wanted a way to see this entire landscape at a glance---which chart types are neighbors (differing by a single feature) and which are far apart.
Not every chart satisfies all of our essential features. Some share many of them, placing them close to Gantt charts without being one, while others share only a few and sit further away. We wanted a way to see this entire landscape at a glance---which chart types are neighbors (differing by a single feature) and which are far apart.

Hasse diagrams~\cite{birkhoff1948lattice, davey2002introduction} offered exactly this. Originally developed to visualize partial orders in mathematics, a Hasse diagram represents elements as nodes and draws edges only between immediate neighbors, omitting redundant connections (most closely related to our usage is Hibbard \etals{}~\cite{hibbard1994lattice} lattice model for data display). 
\begin{wrapfigure}[5]{ri}{0.3\linewidth}
  \vspace{-0.4cm}
    \centering
    \includegraphics[alt={A diamond-shaped Hasse diagram for sets generated by a and b: the set containing both elements is at the top, the two singleton sets are in the middle, and the empty set is at the bottom, with edges between immediate subset relations.},width=1\linewidth]{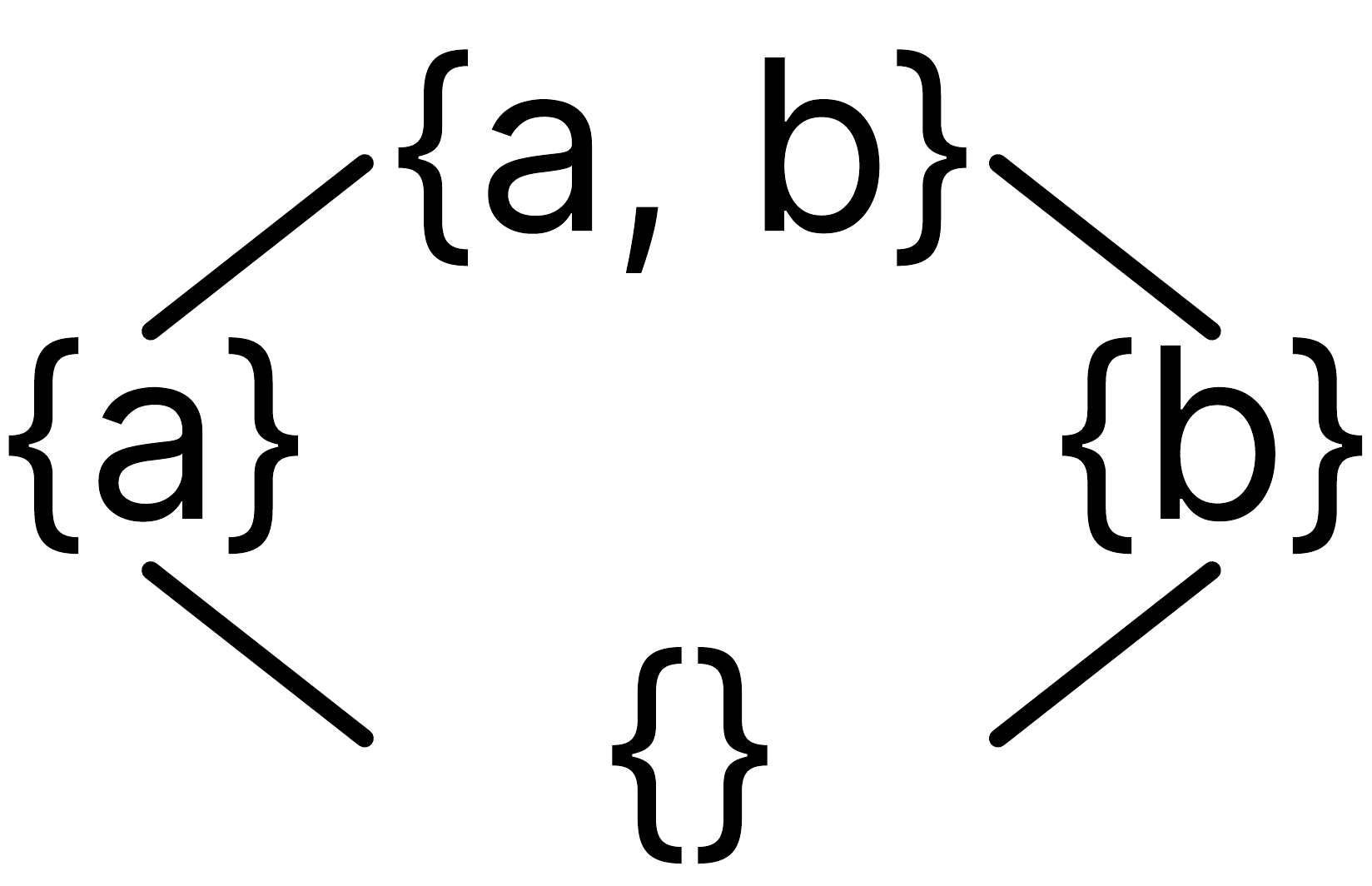}
  \vspace{-0.9cm}
\end{wrapfigure}
We adapted them by letting each node represent a combination of essential features: the top node is the chart type itself (all features present), the bottom node shares none, and the nodes in between represent chart types that satisfy some subset. In settings with many features, alternative set-based visualizations such as UpSet plots~\cite{lex2014upset} may be more appropriate, but Hasse diagrams suit the smaller feature sets here. \autoref{fig:gantt-hasse} shows the resulting diagram for Gantt charts.

\parahead{Merging dependent essential features}
Building the Hasse diagram also prompted us to consolidate our essential features.
We had proposed four, two of which were \essence{Data marks can encode duration} (\textit{Duration}) and \essence{(The chart has a) timeline axis} (\textit{Timeline}). The Hasse diagram required us to populate a node where \textit{Duration} was present but \textit{Timeline} was absent: a chart whose marks encode duration, but that has no timeline axis. We tried to find or imagine such a chart, but encoding duration with data mark lengths inherently requires a temporal reference---you cannot express ``how long'' without some axis against which length is measured.
The node was unpopulable: \textit{Duration} functionally depends on \textit{Timeline}. This led us to consolidate the two into one: \essence{Data marks can encode duration along timelines}.

% \begin{figure}[t]
%     \centering
%     \includegraphics[width=\linewidth]{figures/radar-hasse.pdf}
%     \caption{A Hasse diagram for radar charts and their nearby relations. This representation highlights connections with nearby chart types, such as parallel coordinates or radial line charts.
%     }
%     \label{fig:radar-hasse}
%         \vspace{-2em} 
% \end{figure}

% \begin{figure}[t]
%     \centering
%     \includegraphics[width=\linewidth]{figures/radar-err-hasse.pdf}
%     \caption{
%     Hasse diagrams can be used to identify unnecessary essential features, such as here for radar charts.
%     }
%     \label{fig:radar-err-hasse}
%         \vspace{-2em}
% \end{figure}

\begin{figure}
  \centering
  \includegraphics[alt={Process diagram for defining chart-type boundaries. A vertical bank of curated examples feeds an iterative central workflow. In the upper green phase, Forming candidate features, researchers combine coding examples, expert review, and prior definitions. A central band labeled Defining chart type boundaries links this phase to the lower pink phase, Assessing candidate features, which applies a Hasse diagram, counterexamples, and a flexibility check. Assessment can send the process back to revise the candidates, while a downward arrow leads from the iterative workflow to the output, Functional Definition and Chart Relations. Thus examples inform both formation and testing, and iteration continues until the features support a purpose-specific definition and a map of neighboring chart types.},width=0.87\linewidth]{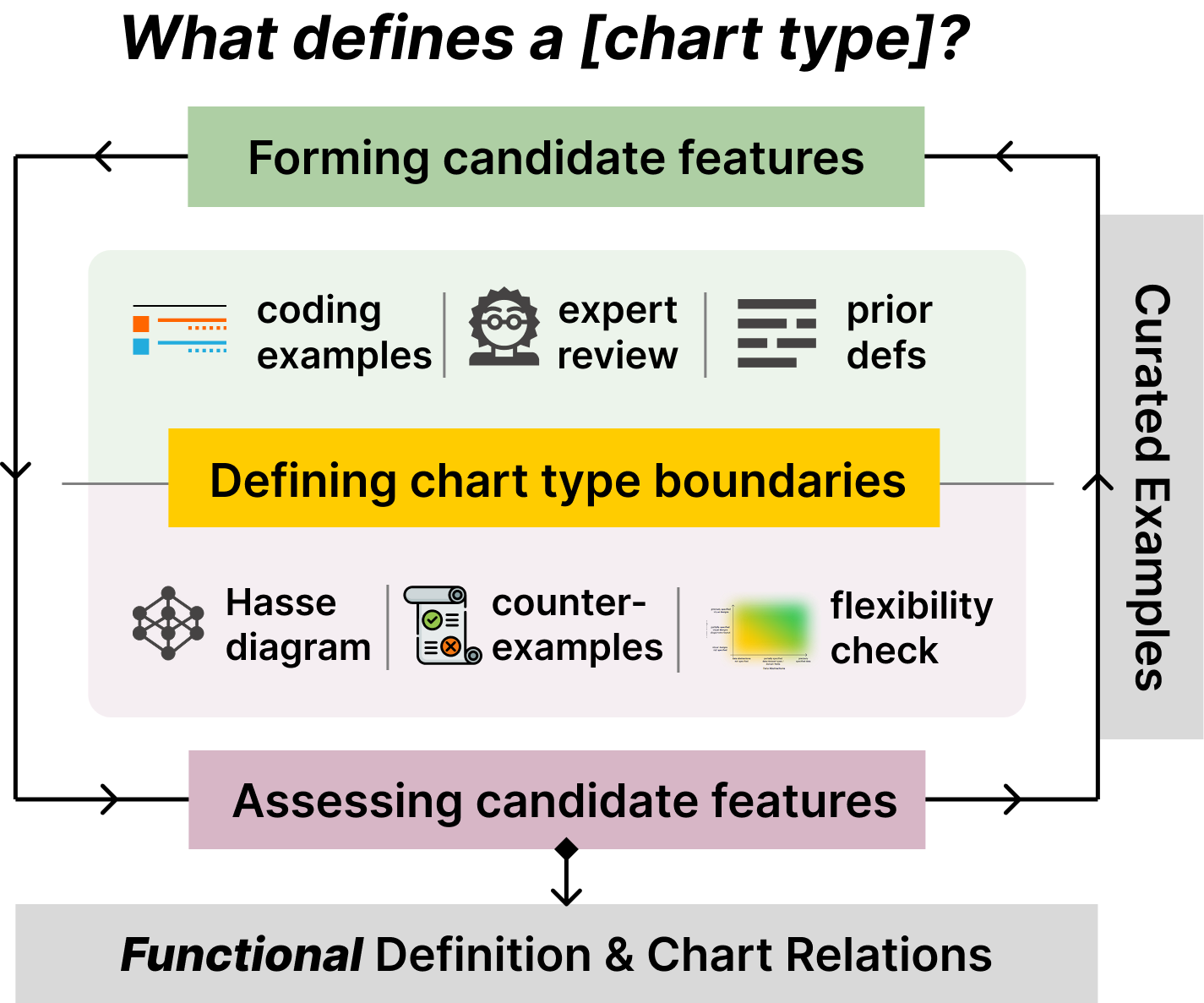}
  \caption{
    An overview of our process for defining chart type boundaries.
    Curated examples feed an iterative loop of forming candidate features
    % (via coding examples, expert review, and axiom revisit)
    and assessing
    which are essential,
    % (via Hasse diagrams, counterexamples, and flexibility checks), 
    yielding functional definitions and an understanding of topological chart relations.
  }
  \label{fig:overview}
  \vspace{-1.5em}
\end{figure}
\subsection{A purpose-built definition}
\label{sec:definition}
% \begin{wrapfigure}[9]{ri}{0.38\linewidth}
%   \vspace{-0.4cm}
%     \centering
%     \includegraphics[width=1\linewidth]{figures/gantt-illustrate.png}
%   \vspace{-0.9cm}
% \end{wrapfigure}
% We summarize our journey and provide an overview in~\autoref{fig:overview}. The previous subsections each resolved a different part of the same problem. Revisiting prior definitions showed that existing accounts of Gantt charts diverged with research purpose. Separating essential from variable features gave us a way to decide which properties belonged to the boundary of the chart type at all. The case examples then refined individual candidates: some broadened overly specific descriptions, some clarified what a feature must be able to encode, and some split apart concepts that had been collapsed under familiar terms. Finally, the Hasse diagram added a structural check, exposing dependencies between features that were not visible in a list alone.
We moved among prior definitions, example-based probes, and Hasse diagram reasoning across multiple rounds of iteration, using each to address different kinds of definitional uncertainty and gradually arriving at a workable definition of Gantt charts. We summarize this in~\autoref{fig:overview}.
After iterating, we arrived at three essential features for Gantt charts:
\begin{enumerate}[nosep, leftmargin=*]
  \item \essence{Data marks can encode duration along timelines}
  \item \essence{Discrete tracks}
  \item \essence{Cross-track dependencies}
\end{enumerate}
Each was shaped through encounter with specific examples that challenged earlier formulations, and each survived both linguistic refinement and structural scrutiny. Together, they define the boundary of the Gantt chart as operationalized for our design space. With that boundary in place, we could turn from the question of \textit{what a Gantt chart is} to the question of \textit{how Gantt charts can vary within that boundary}.

% The resulting definition is illustrated on the right and presented structurally in~\autoref{fig:gantt-hasse}.

% \begin{figure}
%     \centering
%     \includegraphics[width=\linewidth]{figures/gantt-analysis-history.pdf}
%     \caption{The iterative process through which we arrived at our final essences for Gantt charts. Each version was prompted by encounters with examples that challenged the previous formulation.
%     }
%     \label{fig:gantt-analysis-history}
%     \vspace{-2em}
% \end{figure}

\subsection{From definition to design space}
\label{sec:showtime-case}

With that working definition in hand, we could return to our original goal: building a design space for Gantt charts. The essential features now marked the boundary, so the next question concerned the interior: what are all the ways a chart can vary while still being a Gantt chart?

\parahead{Essential features as scaffold}
In \autoref{sec:what-must} we discussed how essential features define the boundary of the design space while variable features fill its interior.
Building the design space made this relationship concrete: each essential feature simultaneously constrains and generates.
\essence{Cross-track dependencies} requires that a Gantt chart encode dependencies between events on different tracks, but within that constraint, many design choices open up. 
% Cross-track dependencies imply discrete tracks, but not vice versa. We kept discrete tracks separate because charts with tracks but no dependencies still form a useful nearby family for the design space. By contrast, charts with duration but no timeline did not, so those features were merged.
Dependencies can be represented as explicit lines between bars, as implicit ordering, or through color encoding; lines can be directed (with arrows) or carry information via thickness.
Likewise, \essence{discrete tracks} constrains the chart to organize events into separate lanes, but leaves open how tracks are ordered, grouped, nested, or labeled.
\essence{Data marks can encode duration along timelines} requires temporal encoding, but leaves open the orientation of the timeline, the scale (linear, logarithmic, or otherwise), and the shape and style of the marks.
% Each essence is simultaneously a wall and a door: it closes off what does not belong and opens a space of possibilities within.
% The variable features that populate these spaces are the dimensions along which Gantt charts differ from one another.

% \begin{figure}
%   \centering
%   \includegraphics[width=1\linewidth]{figures/design-space-clip.png}
%   \caption{An overview of the interactive design space website. It consists of (A) the design space that supports detail-on-demand interaction, (B) a chart gallery where users can select a chart to see how this chart design projects into the design space, and (C) metadata of the chart (source, original name, short description). 
%   % 
%   }
%   \label{fig:design-space}
%   \vspace{-1.5em}
% \end{figure}

\parahead{Scoping the design space with the Hasse diagram}
\label{sec:applyhasse-designspace}
The Hasse diagram (\autoref{fig:gantt-hasse}) also served as a guide for deciding how far beyond the strict definition to look. A strict reading would include only charts satisfying all three features. But the first-level neighbors (charts missing exactly one feature) share much of the same visual and structural vocabulary, and examining them can enrich the design space.

We examined each neighbor and made a deliberate choice.
Charts missing \essence{cross-track dependencies}---such as simple timelines and scheduling displays without inter-event constraints---share enough variable features with Gantt charts to be informative. Their encodings of duration, tracks, and temporal layout transfer directly, and many practitioners would recognize them as Gantt-like.
% Charts missing \essence{discrete tracks}---such as continuous temporal visualizations---similarly offered transferable design choices; we included them too. Charts lacking \essence{duration encoding}---such as certain temporal network diagrams---were different: their variable features (node shape, edge routing, network layout) did not transfer meaningfully. We excluded them.
Charts missing \essence{discrete tracks}, such as continuous temporal visualizations, similarly offered transferable design choices, so we included them. By contrast, charts lacking \essence{duration encoding}---such as certain temporal network diagrams---shared fewer transferable features with Gantt charts: design choices such as node shape, edge routing, and network layout did not meaningfully inform the Gantt design space. We therefore excluded them from it.
Without the Hasse diagram, these decisions would have been ad hoc. The diagram made them principled and traceable.
% We chose to include examples from both of these classes in the design space,
% treating them as \textit{nearby} variants whose design alternatives could enrich our understanding of the Gantt form.

% The third neighbor---charts lacking duration encoding, such as certain temporal network diagrams---was a different case. These charts' variable features (node shape, edge routing, network layout) did not transfer meaningfully to Gantt chart design. Including them would have expanded the space without informing it. We excluded them.

This decision was enabled by the Hasse diagram's structure. Without it, choosing which borderline charts to include would have been ad hoc---a judgment call made in passing and left undocumented.
The diagram made the decision principled and traceable: include neighbors whose variable features transfer, exclude those whose do not.

\parahead{The resulting design space}
% We populated the design space by coding \numInstances{} Gantt chart examples collected from academic papers and practitioner websites (see supplement), identifying variable features under each essential feature through iterative discussion.
Building on the collected examples, we populated the design space by identifying variable features under each essential feature through iterative coding and discussion.
The result is presented as an interactive website
(\href{https://hconhisway.github.io/GanttDesignSpace//}{hconhisway.github.io/GanttDesignSpace}.
The browser uses a detail-on-demand interface: users expand the design space tree to explore variable features, and can select chart examples from a gallery to see how each projects into the space.
Charts satisfying all three essential features are displayed as Gantt charts; nearby variants (those within one hop) are distinguished by graying out the essential feature they lack.

The design space we set out to build was, in the end, shaped by the definition process as much as by the examples themselves.
The essential features determined what the space could contain; the Hasse diagram determined how far beyond the strict definition we should look; and the iterative probing ensured that the boundary was drawn deliberately rather than by default.
This organization also provides a basis for a grammar of Gantt charts: the essential features define constraints that generated charts must satisfy, while the variable features identify the choices such a grammar would need to express.

% This process does not admit a train/test split, because the boundary of ``Gantt chart'' was itself under construction rather than a fixed ground truth to recover. Our safeguards against overfitting were different in kind: a heterogeneous example set that included deliberate edge cases, iterative counterexample-driven revision, and consistency checks via the Hasse diagram.
Drawing that boundary deliberately, however, is not the same as validating it in the way one validates a classifier.
This process does not admit a train/test split, because the boundary of ``Gantt chart'' was itself under construction rather than a fixed ground truth to recover.
Our safeguards against overfitting were different in kind: a heterogeneous example set that included deliberate edge cases, iterative counterexample-driven revision, and consistency checks via the Hasse diagram, \chang{until the definition reached conceptual saturation~\cite{saunders2018saturation} where no new changes emerge from the iteration.}
Defining a chart type is not a classification task with a correct answer waiting to be approximated, it is an inherently subjective process of constructing a boundary for a purpose, and the relevant quality criterion is not accuracy but traceability.

\begin{figure}[t]
  \centering
  \includegraphics[alt={Collection of ten labeled panels, R1 through R10, contrasting radar charts with visually related forms. Green checks identify R1, R2, R4, and R8 as radar charts: R1 is a six-axis multiseries polygon plot; R2 is a four-axis diamond plot using solid and dashed profiles; R4 is a four-axis circular-grid polygon plot; and R8 contains conventional filled radar profiles, including a three-dimensional styled example. Red crosses identify non-radar neighbors: R3 is a radial line plot over ordered network-condition settings; R5 is a radial bar chart by date; R6 contains paired radial line plots over months; R7 contains five annular heatmaps; R9 is a parallel-coordinates plot; and R10 is a circular scatterplot with points of different sizes. The checked examples place values on at least three distinct straight axes with a common origin and connect neighboring values with lines. The crossed examples retain some radial or connecting appearance but violate at least one of those structural conditions.},width=1\linewidth]{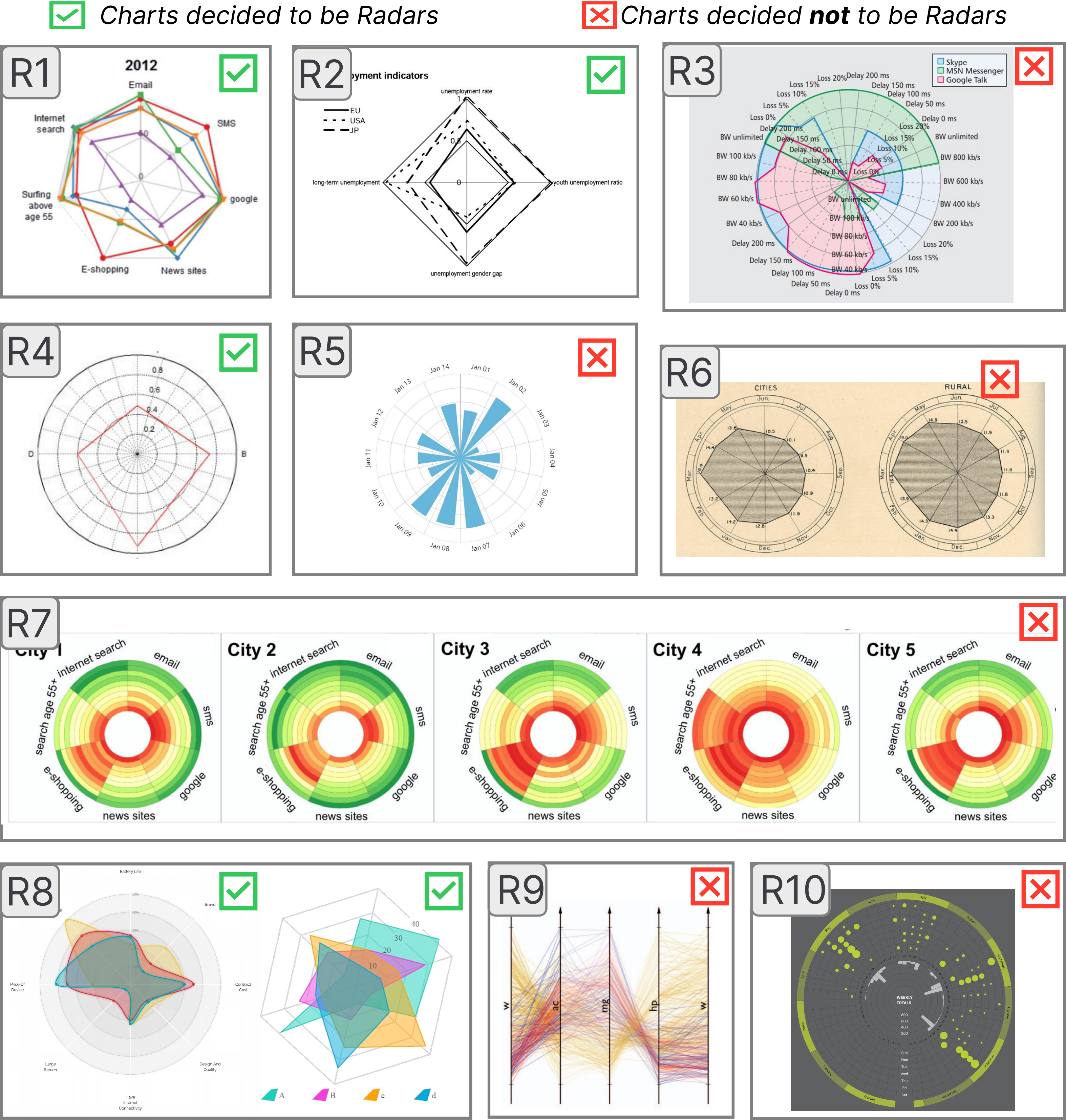}
  \vspace{-1.5em}
  \caption{
  Representative examples of radar charts and related forms. All share some radar-like features (radial axes, multivariate data), but each differs in important ways.
  Sources: 
  R1\&R7~\cite{OfftheRadarAlbo2016}, R2~\cite{Mosley1999}, R3~\cite{Radar2012Chang}, R4~\cite{WANG201742}, R5~\cite{amcharts}, R6~\cite{visualoop}, R8~\cite{visualcinnamon}, R9~\cite{Claessen2011}, R10~\cite{amcharts2}. 
  }
  \label{fig:radarCollect}
  \vspace{-1.5em}
\end{figure}

\section{Beyond Gantts: Radars, Table Cartograms}
\label{sec:other-charts}

The Gantt investigation suggested that chart-type boundaries are not fixed definitions waiting to be found, but boundaries decisions: choices about which features matter, how they relate, and whether and which neighboring forms to treat as part of the type.

To explore what this approach could reveal beyond the Gantt case, we turned to two additional chart types:
% Would the same lesson hold more broadly, beyond Gantt charts? 
% We turned to two additional chart types: 
radar charts, a well-known form with a rich variety of visual variants; and table cartograms, an unusual form whose relationship to more familiar charts is not immediately clear.
The radar case let us see whether defining a structurally different chart type would surface similar kinds of hidden decisions.
The table cartogram case let us ask a different question: what happens when the goal is not to build a definition from scratch, but to understand how an unfamiliar chart type relates to familiar ones?

\subsection{So, what makes a radar chart?}
\label{sec:radar}

Radar charts offered a useful contrast to Gantt charts: visually distinctive and structurally quite different. As in the Gantt case, we began by assembling a collection of examples (\autoref{fig:radarCollect}) from research papers and practitioner websites, deliberately including edge cases.
% ---charts that share radar-like features but might not belong.

% What struck us immediately was not just the visual variety. Some examples used filled areas (\radarRef{6}, \radarRef{7}, \radarRef{8}), others just lines (\radarRef{1}, \radarRef{2}). More revealingly, some charts looked strongly radar-like yet encoded data differently, while others looked nothing like radar charts and still shared most of the same structure.

Porter and Niksiar~\cite{porter2018multidimensional} define radar charts as a \textit{``plot [with] multiple dimensions ($N\geq3$) on a single graphic in the form of closed polygonal profiles.''} Starting from this and from coding our examples, we identified two candidate essential features: that \essence{data is encoded via position} on the axes, and that \essence{data marks are connected}---the lines that link data points into the characteristic polygonal shape. Features like the number of data series, the use of color, shading, and grid lines seemed variable: changing them does not remove the quality of being a radar chart. Removing the connecting lines, on the other hand, would---what remains looks more like a radial bar chart (\radarRef{5}).

% \parahead{Charts that look different, one feature apart}
A first surprise was \radarRef{9}, a parallel coordinates plot. It encodes data via position on axes and connects data marks with lines, so by our first two features it would count as a radar chart. It clearly should not. The missing distinction was axis arrangement: radar charts use \essence{$n (\geq 3)$ equally spaced axes with a common origin}, whereas parallel coordinates use parallel axes. Adding that feature separated the two.

% \parahead{Same shape, different encoding}
Radial line charts (\radarRef{3}, \radarRef{6}) posed the opposite problem. They are often called ``radar charts'' colloquially~\cite{Radar2012Chang}, and at a glance they seem much closer to radar charts than parallel coordinates do: they share the radial layout, the connected line, and often even the polygon-like silhouette. But the important difference is not merely visual. In a radar chart, each axis represents an independent attribute of multivariate data, and the angular position between axes carries no meaning. \chang{In a radial line chart, by contrast, angular position orders observations along a single dimension, often time, much like the x-axis of a Cartesian line chart.}
That distinction forced us to refine one of our candidate essences.
\essence{Data encoded via position} was too permissive: it captured the fact that position mattered, but not how that position was organized. We therefore revised it to \essence{data encoded via position \textbf{on straight axes}}---meaning data points must lie directly on discrete axes, not between them. This excluded radial line charts like \radarRef{3} and \radarRef{6}, \chang{where values are positioned along an ordered angular dimension rather than on separate attribute axes.}
% where values are positioned along a continuous angular dimension.

Alongside these comparisons, we also revisited the wording of the radar essences with the flexibility check introduced earlier (\autoref{fig:ClassifyingDesc}). 
That view made clear that the essences were not equally specific. 
In particular, \essence{data marks are connected} was too loose on both axes: it did not say whether the connection had to be drawn with lines, nor whether any marks or only neighboring ones should be linked.
Tightening it to \essence{neighboring data marks are connected with lines}
\begin{wrapfigure}[7]{ri}{0.55\linewidth}
  \vspace{-0.4cm}
    \centering
    \includegraphics[alt={On axes of visual-design and data-abstraction specificity, an arrow moves from Data marks are connected toward Neighboring data marks are connected with lines, indicating a more explicit commitment on both dimensions.},width=1\linewidth]{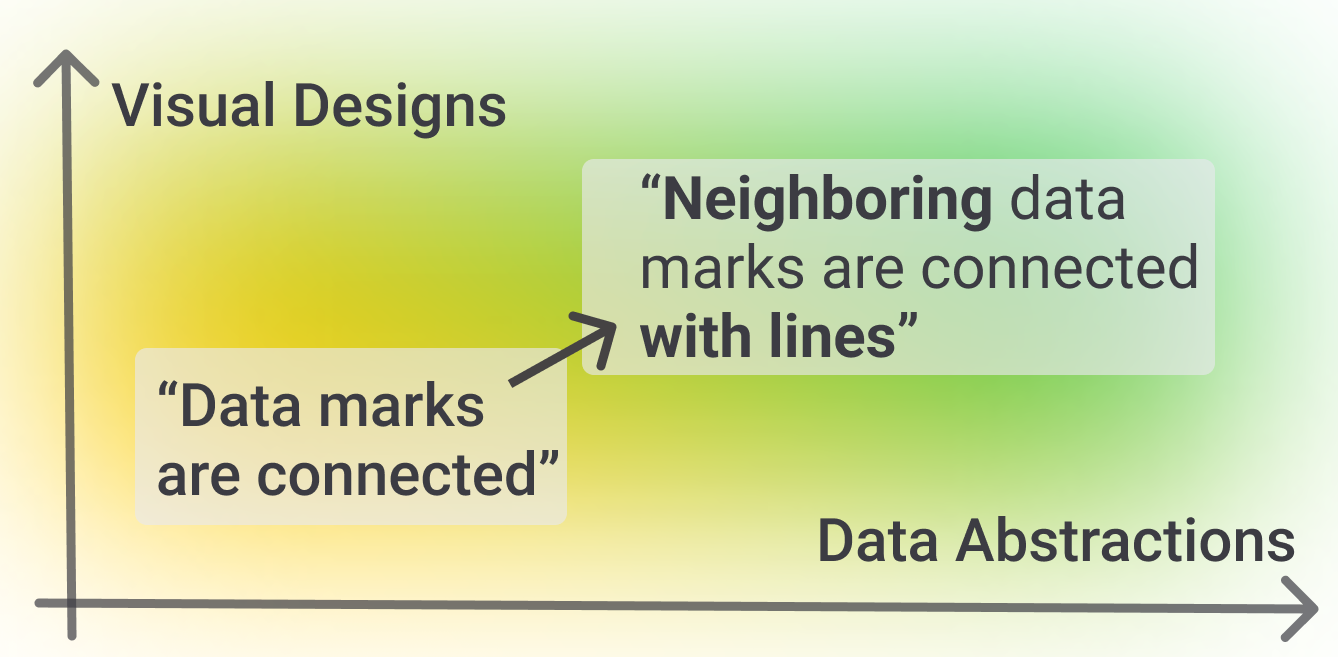}
  \vspace{-0.9cm}
\end{wrapfigure}
made the intended commitment explicit.
The flexibility check therefore helped not only with deciding which charts belonged, but also with deciding how precisely to describe why.

\begin{figure}[t]
    \centering
  \includegraphics[alt={Hasse diagram organizing chart types by three radar-chart features: data encoded by position on straight axes, three or more equally spaced axes with a common origin, and line connectors between neighboring data marks. A blue check means a feature is present, a red cross means it is absent, and each downward edge removes one feature. Radar charts occupy the top with all three. The three immediate neighbors are radial line charts without straight-axis encoding, parallel coordinates without three or more axes sharing a common origin, and flower or radial bar charts without line connectors. Lower nodes retain only line connectors, exemplified by minimal spanning trees; only straight-axis encoding, exemplified by timelines; or only three or more equally spaced axes with a common origin, exemplified by circle charts. The bottom node has none and represents everything else. The diagram makes visually dissimilar parallel coordinates a close structural neighbor while separating visually similar radial forms by the feature they lack.},width=\linewidth]{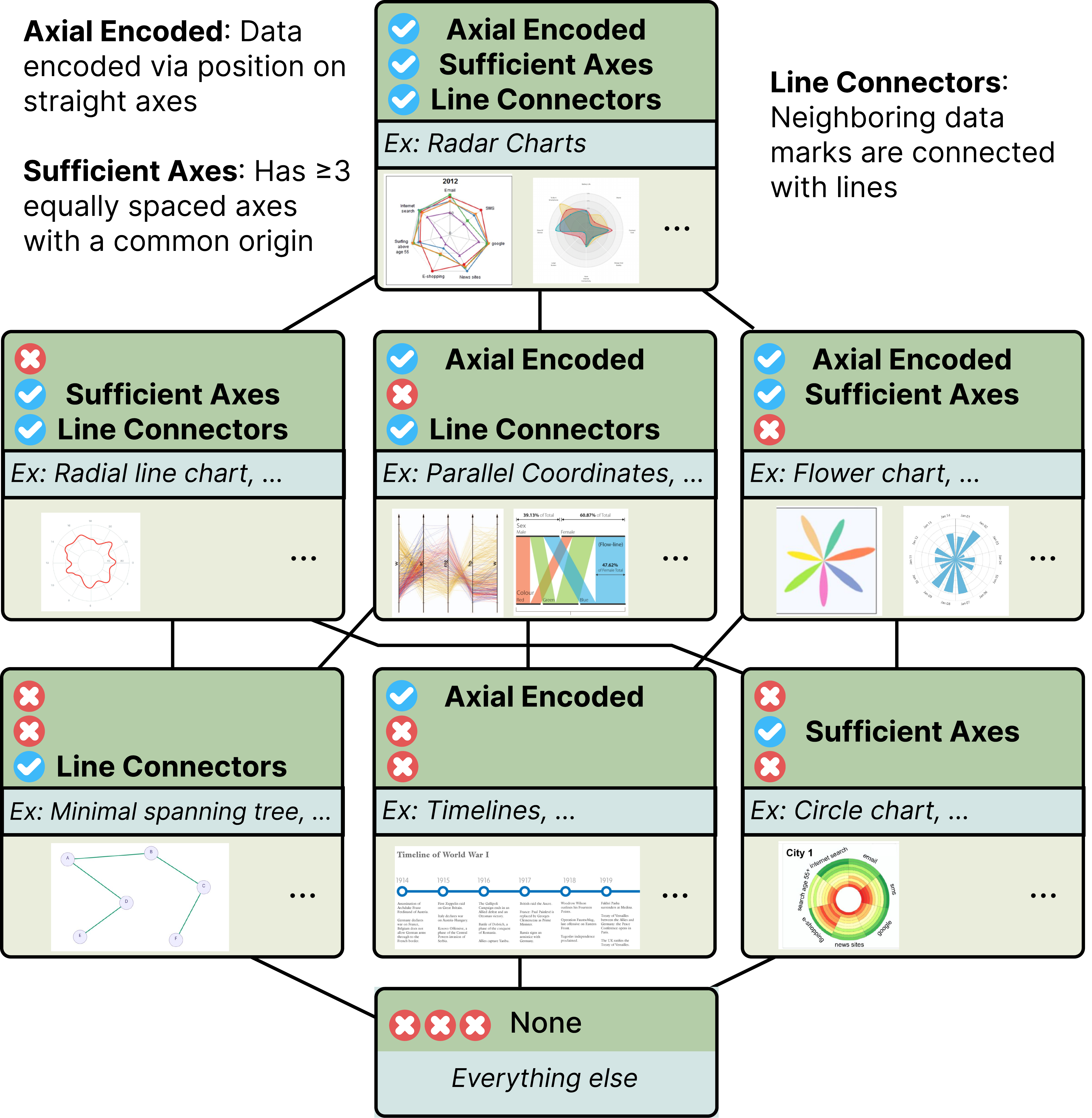}
    \caption{A Hasse diagram for radar charts. Each node represents a combination of essential features. Parallel coordinates, radial line charts, and radial bar charts each sit one feature away---close neighbors that differ in exactly one respect.}
    \label{fig:radar-hasse}
    \vspace{-2em}  
\end{figure}

\parahead{The Radar Hasse}
Using the identified features, we constructed a Hasse diagram for radar charts (\autoref{fig:radar-hasse}).
% , where each node represents a combination of features and each edge removes one feature.
This made the radar chart's closest related chart types easy to see and, in doing so, clarified the boundary of the type.
% Drawing the Hasse diagram for radar charts (\autoref{fig:radar-hasse}) made the boundaries between chart types visible. 
At the top sits the radar chart, defined by all three features. One hop away are parallel coordinates (missing the radial axis arrangement), radial line charts (missing the discrete-axis constraint), and radial bar charts (missing the connecting lines).

The diagram also helped us catch a feature that seemed essential but was not. Early on, we considered whether \essence{data instances must be denoted with categorical colors} should be part of the definition. Many real-world radar charts use color to distinguish data series, so it felt natural. But \radarRef{2} distinguishes series with dashed lines, and \radarRef{4} shows only a single instance. 
% If these are still radar charts---and we thought they were---
\chang{Considering them as radar charts, then categorical color is variable, not essential.} The Hasse diagram made the consequence of including it concrete: an entire class of legitimate radar charts would fall outside the definition.

% \parahead{What radar added to our understanding}
% The radar exercise was lighter than the Gantt investigation, but it sharpened two complementary points. Parallel coordinates provided a useful comparison: although the two chart types look not the same, they are separated by only the layout.
% The radial line chart case showed the converse: charts can look strikingly similar and share a name yet still encode data in importantly different ways. The categorical-color case added a third lesson, namely that a feature that is \textit{prevalent} in a chart type need not \textit{define} it. Together these examples made clear that scope judgments depend on both visual form and data semantics, not on either alone.

\subsection{Situating Table Cartograms}
\label{sec:tacos}
\begin{wrapfigure}[11]{ri}{0.4\linewidth}
  \vspace{-0.4cm}
    \centering
    \includegraphics[alt={A table cartogram composed of a connected grid of irregular quadrilateral cells whose areas and colors encode labeled values. Larger dark-blue cells carry high values such as 90 and 85, while narrow pale cells carry small values.},width=1\linewidth]{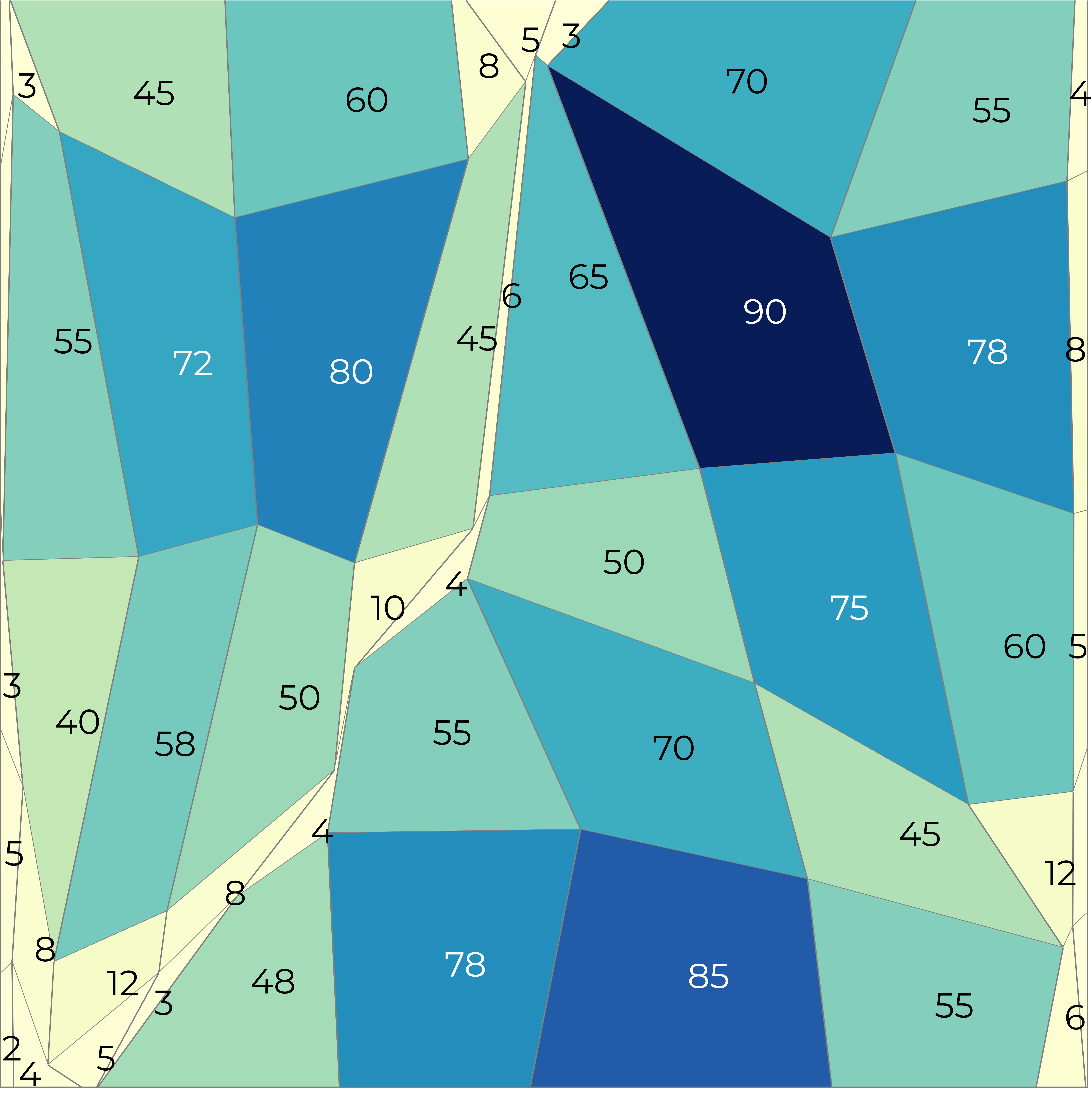}
  \vspace{-0.9cm}
\end{wrapfigure}
Introducing a new chart form is a common occurance in visualization. 
Table cartograms (below) offer a useful example of a related but different task:
not just understanding the chart type itself, as its inventors typically do,
but situating the new form among existing ones. Here the goal is not to build a definition from scratch, but to understand how an unfamiliar chart type relates to more familiar ones.
Broadly, table cartograms are similar to heatmaps that have been ``area-ed'' in addition to shaded. Evans \etal{}~\cite{evans2013table} introduced the form, Hasan \etal{}~\cite{hasan2021putting} improved its construction, and McNutt~\cite{mcnutt2021table} analyzed them via algebraic visualization design.
We took McNutt's description as our starting point, namely that table cartograms combine \essence{Grid-like Planar Topology} (cells do not overlap and their adjacencies are meaningful) with \essence{Accurate Data Embedding} (data is represented accurately as area).

% Suppose though, that this chart form was unfamiliar and that this was our first encounter with it. 
More specifically, table cartograms differ from other cartograms in requiring a connected grid, and they differ from heatmaps in using area as an encoding channel. Using these definitional characteristics, we drew a Hasse diagram to sketch this space of nearby relations.
% As a means to sketch the relationship space we draw a Hasse Diagram for them using the defintional characteristics (an axiomatic approach).

\begin{figure}[t]
    \centering
  \includegraphics[alt={Hasse diagram organizing chart types by three table-cartogram features: meaningful planar topology, connected grid-like quadrilateral marks, and accurate data embedding through area. A blue check means a feature is present, a red cross means it is absent, and each downward edge removes one feature. Table cartograms occupy the top with all three. Immediate neighbors missing one feature are treemaps or mosaics without planar topology, most cartograms without grid-like quadrilaterals, and heatmaps or tables without area embedding. Lower nodes retain only accurate area embedding, exemplified by Dorling cartograms and Voronoi treemaps; only planar topology, exemplified by choropleths, other maps, and grids; or only grid-like quadrilaterals, exemplified by geometric designs. The bottom node has none and represents everything else. Separating planar topology from grid-like quadrilaterals keeps treemaps distinct from table cartograms and clarifies which neighboring forms share particular structural properties.},width=\linewidth]{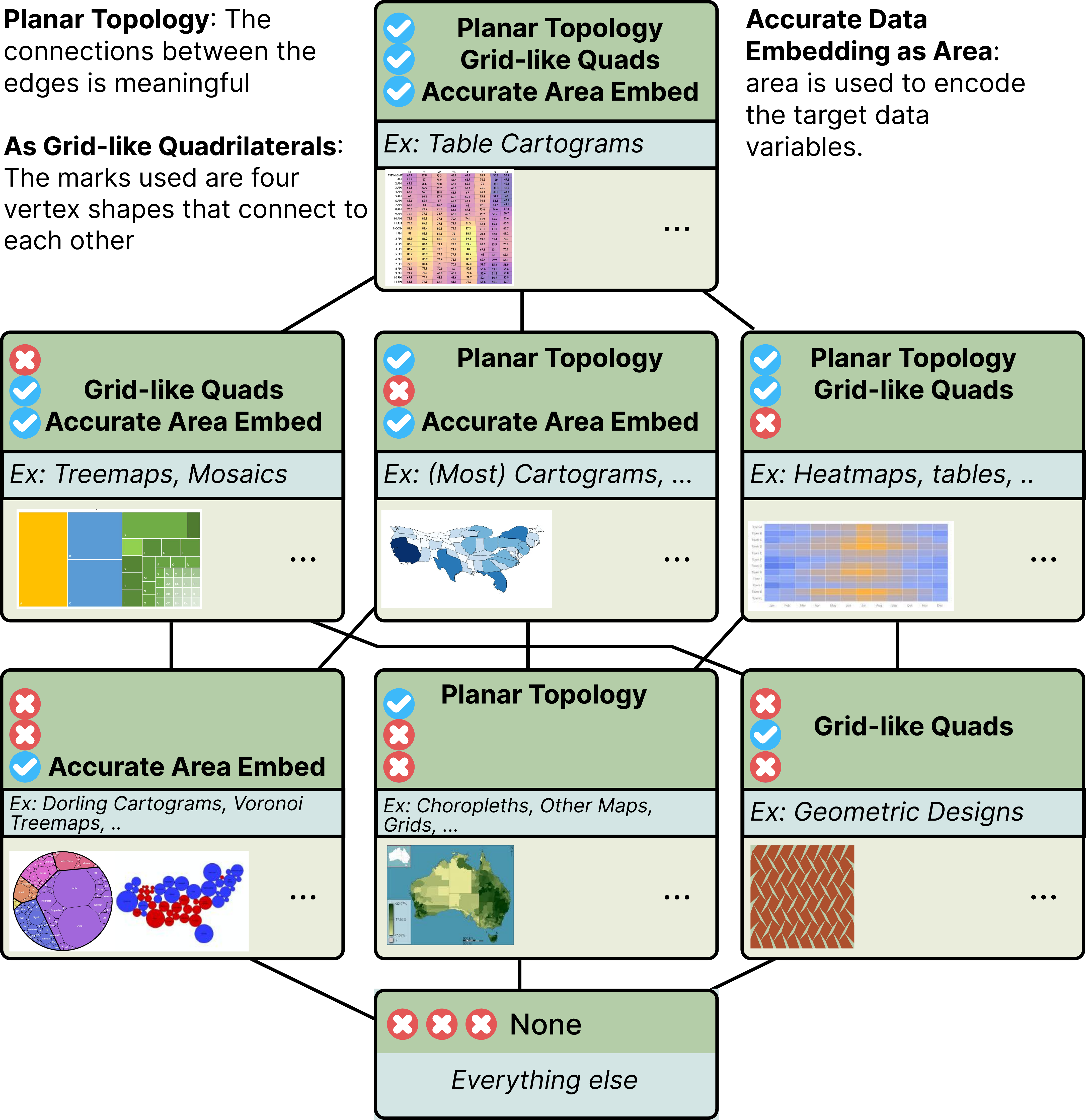}
    \caption{A Hasse diagram for table cartograms and its nearby relations. }
    \label{fig:taco-hasse}
    \vspace{-1.5em}
\end{figure}

In doing so, we found the initial groupings too coarse, by treating \essence{Grid-like Planar Topology} as a single feature, it grouped treemaps with table cartograms, even though treemaps share the planar subdivision but not the connected grid of quadrilateral cells that table cartograms require.
We therefore split \essence{Grid-like Planar Topology} into \essence{grid-like quadrilaterals} and \essence{planar topology}. The resulting diagram (\autoref{fig:taco-hasse}) materializes that sketched definition and makes explicit its relations to nearby forms.

McNutt~\cite{mcnutt2021table} analyzes table cartograms using a generic visualization task model, while noting that a cartogram-centric one, such as that of Nusrat \etal{}~\cite{nusrat15CartogramTax}, could also be useful. 
% The diagram 
The Hasse diagram helps explain why a generic model is defensible: many of the nearest neighbors are generic information visualizations (\eg{} treemaps and heatmaps) rather than geography-laden cartograms. 
% So while the analysis provides no answers on its own, it enriches the context in which such an analysis can be interpreted. 
It can also help organize related work, situating a reader's mental model of the new chart form within more familiar ones.
% A natural extension would be to connect McNutt's Algebraic Visualization Analysis~\cite{mcnutt2021table} with our Hasse diagrams so that findings in one cell could be related to others. 
This raises interesting questions. For instance, table cartograms share grid-structured data with heatmaps but not with geographic cartograms, suggesting that some guidelines may transfer to the former but not the latter. 
% We leave that more explicit account of transfer to future work.

The table cartogram case led us to a complementary realization: defining a chart-type boundary requires situating a chart among its neighbors. In that sense, situating an unfamiliar form and defining a familiar one are not fundamentally different activities, but differently motivated versions of the same boundary work. It also highlighted a question of granularity: a feature that is sufficient for describing a chart may still be too coarse for reasoning about its nearby relations, and once those relations are made explicit, it becomes clearer which task models and design guidance are likely to transfer.

% \begin{boxH}
% \emph{Relating new chart types to familiar ones is a difficult task to do systematically, our framework offers a means to explicate the studied and described relations. }
% \end{boxH}

\section{Summary: On Defining Chart Type Boundaries}

% Taken together, our cases---alongside patterns observed in the broader visualization literature---suggest three insights about what is at stake when defining chart types.
Our cases suggest three related insights about what is at stake when defining chart types. These insights move from why definitions differ, to what making boundaries explicit reveals, to why those scope choices matter for interpreting findings.

% \parahead{Definitions Diverge for Functional Reasons}
First, our cases suggest that divergent chart-type definitions are not failed attempts to converge on a single correct answer, but rather boundary choices made to support a particular research aim. Rosch~\cite{rosch1975cognitive} showed that categories are often organized around prototypes rather than strict necessary-and-sufficient conditions, which helps explain why different researchers may begin from different exemplars of the ``same'' chart. 
Our cases suggest that chart-type definitions in research are better understood as purposeful endeavor to make a vague concept more precise. 
% precisifications
% of a vague concept. 
In Carnap's terms, they are \textit{explications}~\cite{carnap1950logical}: more precise formulations constructed for particular analytic aims. Understanding these definitions as purpose-built explications helps explain 
%  helps explain
why prior definitions of Gantt charts diverged without any one of them simply being wrong. They were not competing for ownership of the one true Gantt chart, but defining related objects suited to different research purposes. As Goodman~\cite{goodman1978ways} argued more broadly, multiple categorizations can be valid when they serve different ends.
% When read through this lens, implicit scope choices across the visualization literature reveal the same pattern: Felix \etal{}~\cite{felix2017taking} reframe ``word clouds'' as ``keyword summaries'' to suit an analytic rather than aesthetic purpose; Brehmer \etal{}~\cite{brehmer2016timelines} scope ``timelines'' for expressive storytelling, encompassing Gantt-like forms under a framing that project-management research would not share; Kong \etal{}~\cite{kong2010perceptual} and Vernier \etal{}~\cite{vernier2020quantitative} scope ``treemap'' to rectangular and time-dependent variants, respectively, for different evaluative ends. In each case, the definitional boundary shifted with the functional goal.

Thus, the practical consequence is that asking ``which definition is correct?'' might be the wrong question; the productive question is ``what is this definition for?''

% \parahead{Boundary Work Reveals Hidden Structure}
% Once that question is foregrounded, 
% Second, our cases suggest that making chart-type boundaries explicit does more than settle edge cases: it reveals hidden structure in the descriptive vocabulary itself.
% Second, our cases suggest that making the boundary explicit becomes analytically useful, because it reveals the assumptions built into the definition itself.
A second insight is that defining a chart type does more than settle edge cases---it enforces stress-testing the descriptive vocabulary researchers use for chart types and exposes assumptions that remain invisible when categories are taken for granted. Before boundary work begins, features appear independent, visual similarity appears to track structural similarity, and feature descriptions appear to be at the right level of specificity. All three assumptions broke down in our cases. In the Gantt case, features that appeared independent turned out to be entangled: duration encoding presupposed a timeline axis, so the two could not be specified separately. In the radar case, visual similarity turned out to be a poor proxy for structural equivalence: radial line charts look like radar charts but encode data along a fundamentally different dimension, while parallel coordinates look different from radar charts,
yet share all but one structural commitment. 
% And 
% feature descriptions turned out to carry implicit granularity choices: a characterization sufficient for describing a chart proved too coarse for distinguishing it from its neighbors, requiring decomposition before the Hasse diagram could do useful work (table cartogram). 
For table cartograms, a feature description that was adequate for describing the chart proved too coarse for distinguishing it from nearby forms: \essence{Grid-like Planar Topology} had to be decomposed before the Hasse diagram could separate them from treemaps and other neighbors.

Drawing a boundary exposed dependencies among features, ambiguities in their meaning, and choices about the level at which a chart type should be described. The result was not just a cleaner definition, but a clearer account of the conceptual structure that the definition imposed.

A third insight is that scope choices shape the reach of generalization.
% Those assumptions matter for more than description. 
Once boundary work clarifies what kind of chart is actually being studied, it also clarifies how far the resulting claims can be generalized.
Scope choices therefore do not merely determine what belongs to a chart type; they also determine which examples count as evidence, which neighboring forms are relevant comparisons, and which findings should be expected to transfer.
% This 
This relationship between scope and generalization became especially clear in the Gantt case, where some closely related forms supported meaningful comparison while others did not. What mattered was not simple visual resemblance, but whether the differing features were relevant to the design question at hand. The radar case made the complementary point: charts that share a name may encode data in importantly different ways, so findings about one need not automatically be expected to apply to the other.

Across these cases, the lesson is that chart-type-centered findings attach not to names alone, but to the particular variant of a chart that a study includes and excludes.
Making chart-type boundaries visible is therefore not merely terminological cleanup; it is part of making chart-type-centered research interpretable, comparable, and cumulative.
Concretely, this means a reader can tell which variant of a chart a study's findings describe, whether two studies using the same chart-type label examined the same object, and how far a result should be expected to transfer to neighboring forms.

\section{Discussion}
\label{sec:disco}
% The broader implication of this work is less about Gantt charts alone than about what chart-type names do in visualization research. Design spaces, perceptual studies, grammars, and comparisons all rely on familiar labels such as ``Gantt chart'' or ``treemap,'' yet those labels are often treated as if their boundaries were already settled. Our investigation suggests a different view: a chart definition is not merely a name applied to a preexisting object, but a research construction that helps specify what object is being studied. We discuss four implications of this view.
Chart-type names are easy to invoke and difficult to delimit. Design spaces, perceptual studies, grammars, and comparisons often rely on familiar labels such as ``Gantt chart'' or ``treemap,'' yet those labels are often treated as if their boundaries were already settled. Our investigation suggests a different view: a chart definition is not merely a name applied to a preexisting object, but a research construction that helps specify what object is being studied.

% \parahead{Definitions are research constructions}
\parahead{Definitions are shaped by purpose}
We did not discover the true definition of a Gantt chart. We constructed a working one adequate to a particular goal, and in doing discovered that the question ``what is a Gantt chart?'', in practice, is fundamentally tied to the question ``for what purpose?'' This relationship does not make chart definitions arbitrary. It means they should be judged by whether they draw (and explicate) a reasonable boundary for the current task at hand.
% .and make that boundary inspectable to others.
\chang{This purpose enters throughout the decisions in the process: which examples to collect, which borderline cases to test, and how broadly to phrase each essential feature. Because we aimed to build a Gantt chart design space, we favored a scope broad enough to capture useful design variation yet narrow enough that the included cases shared a meaningful design vocabulary---a goal that shaped our corpus, our handling of edge cases such as spring-based and temporal-network-like Gantts, and our decisions about feature granularity and wording.}

The literature suggests that this situation is common rather than exceptional. 
Felix \etal{}~\cite{felix2017taking} deliberately reframe their object of study from ``word clouds''---a name carrying aesthetic and emotional connotations---to ``keyword summaries,'' because their functional goal of studying exploratory data analysis demanded a different scope than work treating word clouds as playful artifacts. The visual form did not change; the boundary did. Brehmer \etal{}~\cite{brehmer2016timelines} scope ``timelines'' broadly enough to encompass Gantt-like forms under a storytelling framing that project-management research would not share. Kong \etal{}~\cite{kong2010perceptual} and Vernier \etal{}~\cite{vernier2020quantitative} study differently scoped treemap variants for different evaluative ends. Historical drift in Gantt charts makes the same point from another angle: in Gantt's original use of the form~\cite{gantt1903}, there were no encodings of cross-track dependencies (though Adamiecki may have contemporaneously developed timeline charts with cross-track dependencies~\cite{Marsh1975Harmonogram}), whereas such dependencies are now standard in widely-used tools. What changes in these cases is not only terminology, but the research object stabilized by that terminology for a given use. Divergent chart definitions are therefore often better understood as different operationalizations than as failed attempts to identify a single natural kind.

\parahead{Boundary work is diagnostic}
The most important payoff of making a boundary explicit is not merely that edge cases get resolved; it is that the attempt exposes hidden structure in the vocabulary used to describe a chart type. In our cases, supposedly independent features turned out to be entangled, apparent visual similarity turned out to mask different encodings, and familiar feature descriptions turned out to be too coarse to distinguish nearby forms. 
% The distince between essential and variable features and the Hasse diagrams were valuable less because they resolved these issues automatically than because they made them visible and debatable.

Similar diagnostic payoffs appear elsewhere in the literature. Skau and Kosara~\cite{skau2016arcs} show that pie charts are not mainly read through angle, exposing a granularity gap in how the chart is often described. Hearst \etal{}~\cite{hearst2019evaluation} show that semantically grouped word-cloud layouts outperform standard Wordles, suggesting that spatial arrangement---often treated as part of the form's identity---may instead be a variable feature. Together these cases suggest that boundary work is not only classificatory labor; it is also a way to stress-test descriptive concepts and revealing where they are too coarse, too independent, or too casually inherited.

This diagnostic role also mattered practically in our own process. When coding examples, the question of whether a chart \textit{was} a Gantt chart shifted from a holistic impression to a revisable argument about features. Likewise, in the Gantt case, the Hasse diagram helped us reason about which nearby forms were worth preserving for design space construction and which were too distant to be useful.

\parahead{Scope conditions what can be claimed}
Once chart definitions are understood as research constructions and boundary work as diagnostic, a further implication follows: scope is part of a study's evidential basis. Scope determines which examples count as evidence, which neighboring forms are relevant comparisons, and how far findings can reasonably be generalized. In our Gantt work, we included two nearby forms in the design space browser because our goal---producing a design space---favored preserving charts with potentially useful design features. We excluded neighbors lacking duration encoding because they were functionally too distant. That scoping decision, in turn, limits what our resulting design space can claim: it applies to the family of duration-encoding Gantt variants we treated as evidence, not to all nearby scheduling or timeline-like forms.

One might even argue that scope is one expression of an inevitable interpretive bias. This need not be read as a criticism. Researchers may use the same chart-type name while carrying different, contextually situated understandings of what that name denotes, shaped by task, domain, disciplinary background, and prior examples. The aim is therefore not to eliminate such bias and recover a perfectly neutral definition, but to keep it in view: to make scope decisions explicit so that design, analysis, comparison, and evaluation proceed with awareness of the assumptions built into them. This orientation resonates with feminist epistemological arguments for surfacing the situatedness of knowledge rather than assuming neutrality~\cite{d2016feminist, akbaba2024entanglements}.
\chang{Practically, researchers can state what a definition is for, and then use these tools to investigate and document how that purpose shaped inclusion and exclusion decisions, and make the situated commitments behind the boundary available for inspection and revision.}

The same dependency is visible in prior research. 
% Bertini \etal{}~\cite{bertini2020shouldn} note that Cleveland and McGill's channel-effectiveness rankings do not hold when the task shifts from individual value comparison to trend detection. Our radar case makes the same point from the opposite direction: charts that share a name may still encode data in importantly different ways, so findings about one need not apply to the other. 
Bertini \etal{}~\cite{bertini2020shouldn} note that Cleveland and McGill's channel-effectiveness rankings do not hold when the task shifts from individual value comparison to trend detection---the same channels, measured differently, yield different results. Our radar case illustrates a complementary vulnerability: charts that share a name may still encode data in importantly different ways, so findings about one variant need not apply to another. In both cases, what changes is a condition that is easy to overlook but consequential for generalization.
Chart-type-centered findings therefore attach not to names alone, but to the particular variant of a chart that a study includes and excludes. Reporting scope is not terminological housekeeping; it is part of stating the conditions under which a result should be interpreted, compared, or transferred.

\chang{

\parahead{Features are scoped commitments}
We borrow the essential-variable distinction as a pragmatic vocabulary for scope reasoning, rather than as a claim about context-independent essences of chart types.  Discussions of essentialism have long noted that what counts as essential can depend on how an object is being considered~\cite{sep-essential-accidental}, and this dependence is central to how we use the distinction: a feature is essential or variable only relative to a scoped object of inquiry and the purpose behind it.

This matters for multi-view and composite visualizations. A visualization may contain, combine, or coordinate multiple chart-like parts without the whole display being an instance of each chart type it contains. In our Gantt case study, we therefore analyzed separable Gantt portions of multi-view systems as the relevant chart instances, rather than labeling the entire systems as Gantt charts; that decomposition was itself a scope decision, not a neutral fact about the display. The same issue applies to composite charts more broadly: a study of a map--matrix combination such as MapTrix~\cite{yang2016many} might focus on its map portion if the object of inquiry is maps, while a study of map-based composites would need to treat the relations among parts as part of what defines the object of inquiry.

This scope decision also determines where feature claims attach. When the object of inquiry is a single chart type, features are attributed to that chart; when it is a composite, features may include relations among parts, such as coordination, alignment, or shared data objects. Our cases focus on chart-type definitions rather than whole multi-view systems; applying the same reasoning to multi-view and composite visualizations may require additional considerations for composition, coordination, and interaction.
% What should be avoided is shifting levels mid-analysis: collecting duration encoding from one part, dependencies from another, and treating their union as the definition of the whole. 
% The issue is therefore not to supply a universal definition of ``component,'' but to make explicit the level at which the definition is being constructed.

}

% \parahead{Where this reasoning may extend}
\parahead{Where this boundary reasoning may extend}
Although we developed these ideas through chart types, the same style of reasoning may be useful for other contested visualization concepts. Annotations, for example, still lack a settled definition~\cite{dilshadur2025}, and the distinctions developed here may help articulate where one annotation type ends and another begins. 
% Composite and multi-view visualizations pose a harder but promising extension: the choice to decompose a system into constituent charts is itself a scope decision, not a neutral preprocessing step. 

\chang{At the same time, our tools are unlikely to help equally in all settings. They are most valuable when chart-type boundaries bear directly on the research claim, such as in design spaces, grammars, taxonomies, comparative studies, or perceptual experiments. In these settings, the time and cognitive cost of making scope decisions explicit is often outweighed by the benefit of clarifying what was studied, what was excluded, and how far findings might transfer. A lighter treatment may suffice when the object of study is too unsettled to support stable features, or when a chart name is used as a local design reference in a design study rather than as the basis for generalization. The appropriate level of rigor therefore depends on the genre and stakes of the work.
% our aim is diagnostic rather than prescriptive, supporting reflection rather than mandating a single procedure or definition.
}
% At the same time, our tools are unlikely to help equally in all settings. They are probably less useful when the object of study is too unsettled to support stable features at all, or when the target is a one-off bespoke design rather than a recurring form.

\chang{
\parahead{Feature abstractions trade nuance for simplicity}
Our use of binary, equally weighted features is a deliberate simplification. We treat features as essential or variable, and as present or absent in the Hasse diagrams, not because visualization features are intrinsically binary, but because this abstraction keeps the resulting structure simple and inspectable. 
% Some features can be continuous: such as map-based visualizations can have different levels of approximation in shapes~\cite{Heilmann2004RecMap}. Features may also differ in their influence: some strongly shape a chart's identity, while others matter only locally. 
Features can also be continuous: some parameters interpolate between designs~\cite{schulz2015preset}, while others weight conflicting requirements that cannot all be satisfied at once, such as map-based visualizations can have different aspects of approximation in shapes~\cite{Heilmann2004RecMap}.
% Feature-vector approaches to visualization similarity, such as Li \etal{}'s phylogenetic trees for hierarchical visualization designs, offer one way to reason about such unequal feature influence~\cite{li2015exploring}. 
Understanding such unequal feature influence would require vectoring visualization designs  such as in Li \etal{}'s phylogenetic trees for hierarchical visualization designs~\cite{li2015exploring}.
Estimating the feature weights or deriving feature vectors would also require empirical evidence base from a visualization corpus; for example, this would require a large-scale study of how people perceive the importance of visual features. Such extensions could support broader taxonomy-building efforts, but would add complexity beyond the goal of providing a lightweight structure for making boundary decisions visible.

We likewise simplify the approach by treating features as atomic, though a feature can carry internal structure of its own. The flexibility check surfaces part of this---phrasings differ in the direction and coverage of what they constrain---and related conditions can be folded into a single feature rather than spread across several, which is a decision that should be made based on the analytic purpose and whether the resulting internal distinctions are useful for the task at hand.
}

\section{Conclusion}

% \paraheadd{}hard conclusion

% The larger lesson is that definitional work should not be treated as mere preamble to visualization research. In many chart-type-centered projects, it is part of the research itself, because it determines what counts as the object of inquiry, what counts as evidence about it, and therefore how design claims, grammars, and evaluations should apply.

Defining a chart type is not preamble to visualization research---it is part of the research itself. Every design space, grammar, perceptual study, and comparison that invokes a chart-type name depends on a scope: a commitment to what that name means for the purposes at hand. That commitment determines which examples count as evidence, which variations matter, and how far findings can be expected to generalize. When it remains implicit, so do these consequences, and the work built on it becomes harder to interpret, compare, or extend.

The tools we assembled---the essential--variable distinction, Hasse diagrams, flexibility checks---are not instruments for discovering the ``true'' boundary of a chart type. They are instruments for making a constructed boundary visible: for turning private intuitions into shared, inspectable, revisable artifacts. Their value lies not in the definitions they produce but in the decisions they surface. \chang{Neither should they be seen as prescriptive; they are meant to be applied flexibly, allowing users to derive their own definitions.}

We do not argue for universal definitions, nor that every study must undertake the kind of sustained investigation we describe with Gantt charts here. We argue that every chart-type-centered study already draws a boundary, whether or not it is explicit. We began by asking what makes a Gantt chart. We end with a more practical lesson: when chart-type-centered research runs into questions of scope, the solution is not to find the one true definition, but a way to reason toward functional definitions for the purposes at hand.

\section*{Acknowledgements}
This work was supported by the United States Department of Energy through award DE-SC0024635 and the United States National Science Foundation through award \#2402719. We thank Sayef Azad Sakin for valuable discussions on Gantt charts. We are also grateful to Connor Scully-Allison, Tingying He, Md Dilshadur Rahman, and members of the Scientific Computing and Imaging Institute for their valuable feedback and advice. We also wish to thank the anonymous reviewers for their thoughtful suggestions.
% Omitted for anonymous review.

\section*{Supplementary Material}
We include the following supplemental materials:

\begin{enumerate}
  \item A file containing page-sized versions of our figures \chang{plus a screenshot of the interactive website of Gantt design space.}
  \item A file containing our radar examples and coding results.
  \item A file containing our Gantt chart collection list. Code and files necessary to build and run our Gantt chart design space browser as well as a short video showing the browser's use. The browser itself is available at \href{https://hconhisway.github.io/GanttDesignSpace//}{hconhisway.github.io/GanttDesignSpace}.
\end{enumerate}

The supplemental materials are available at \osf{}. 

% bibtex
% \bibliographystyle{abbrv-doi-narrow}
\bibliographystyle{abbrv-doi-hyperref}
\bibliography{EDSA}       

% biblatex with biber
% \printbibliography                

% \input{appendix}

\end{document}